\documentclass[pra,reprint,amssymb]{revtex4-1}%

\usepackage{amsmath,amssymb}
\usepackage{revsymb4-1}
\usepackage[cmyk]{xcolor}
\usepackage[%
	colorlinks=true,%
	linkcolor=blue,%
	citecolor=blue,%
	urlcolor=blue
	]{hyperref}%
\usepackage{graphicx}%
\usepackage{here}
\usepackage{dcolumn}%
\usepackage{bm}%
\usepackage{revsymb}
\usepackage{braket}
\usepackage{layouts} 
\usepackage{color}
 
\def\equationautorefname~#1\null{\textrm{~(#1)\;}\null}
\def\figureautorefname~#1\null{Fig.~#1\null}

\begin{document}

\title{Density and spin modes in imbalanced normal Fermi gases \\ from collisionless to hydrodynamic regime}
\author{Masato Narushima, Shohei Watabe, and Tetsuro Nikuni}
\begin{abstract}
      We study mass and population imbalance effect on density (in-phase) and spin (out-of-phase) collective modes in a two-component normal Fermi gas. By calculating eigenmodes of the linearized Boltzmann equation as well as the density/spin dynamic structure factor, we show that mass and population imbalance effects offer a variety of collective mode crossover behaviors from collisionless to hydrodynamic regimes. The mass imbalance effect shifts the crossover regime to the higher-temperature, and a significant peak of the spin dynamic structure factor emerges only in the collisionless regime. This is in contrast to the case of mass and population balanced normal Fermi gases, where the spin dynamic response is always absent. Although the population imbalance effect does not shift the crossover regime, the spin dynamic structure factor survives both in the collisionless and hydrodynamic regimes. 
      \end{abstract}
 
\affiliation{Tokyo University of Science, 1-3 Kagurazaka, Shinjuku-ku, Tokyo, 162-9601, Japan}
\maketitle


\section{Introduction}

Collective modes in quantum many-body systems have received much attention because of observation of many-body quantum statistical nature, not only in superfluid states~\cite{Kinast2004,Bartenstein2004,Altmeyer2007,Joseph2007,Wright2007,Tey2013,Sidorenkov2013} but also in normal states~\cite{Gensemer2001,DeMarco2002,Wright2007}. 
In a Fermi system, collective modes in two regimes---the collisionless regime and the hydrodynamic regime---have been a central issue in liquid helium $^{3}$He~\cite{Landau1957,Abel1966}, neutron stars~\cite{Nitsch1972,Kouvaris2009} as well as quantum gases~\cite{Vichi2000,Bartenstein2004,Altmeyer2007,Wright2007}. 
In particular, the zero sound---the in-phase (density) sound mode in a collisionless regime---and the first sound---the density sound mode in a hydrodynamic regime--- have been intensively and extensively investigated in two-component simple normal Fermi systems~\cite{Landau1957,Khalatnikov1958,Nitsch1972,Larionov1999,Watabe2009,Watabe2010}, dipolar Fermi gases~\cite{Ronen2010}, polarized Fermi-liquid films~\cite{Li2013}, as well as Fermi gases with arbitrary spin~\cite{Yip1999}. 

Ultracold quantum gases, in particular, offer a prominent playground to selectively study mass and population imbalance effects.  
Fermi-Fermi mixtures, $^{6}$Li and $^{40}$K, are realized and their interaction strength are controlled by Feshbach resonance~\cite{Trenkwalder2011,Spiegelhalder2009,Tiecke2010}. 
Furthermore, not only have the density mode but also the spin mode been also a central issue to study in this field. 
Possibility of the out-of-phase (spin) modes has been theoretically discussed at unitarity~\cite{Stringari2009}, and normal polarized Fermi gases~\cite{Dahal2008,Recati2010}; spin drag~\cite{Duine2011}, spin-seebeck effect~\cite{Wong2012}, as well as spin diffusion~\cite{Bruun2011} also have been explored. 
Universal nature of spin transport phenomena~\cite{Sommer2011,Koschorreck2013}, and spin transport in polaronic Fermi gases~\cite{Ariel2011} have been experimentally studied, where the spin-diffusion is suppressed in the low-temperature regime~\cite{Sommer2011,Ariel2011}. 
However, mass and population imbalance effect on the collisionless and hydrodynamic density modes as well as on those spin modes is an open issue at finite temperatures. 

In this paper, we study mass and population imbalance effects on the in-phase and out-of-phase collective modes in a two-component normal Fermi gas, focusing on the crossover from the low-temperature collisionless regime to the high-temprature hydrodynamic regime. 
We find that mass and population imbalance effects not only shift eigenfrequency of collective modes qualitatively, but also drastically change the spin mode properties. 
In mass imbalanced gases, the crossover regime is shifted to the higher-temperature, and only in the collisionless regime, a significant peak emerges in the spin response function. 
This is in stark contrast to the case of balanced normal Fermi gases, where the spin dynamic structure factor is absent at all temperatures. 
Although population imbalance effect does not shift the crossover regime, the spin response function survives both in the collisionless and hydrodynamic regimes.

\section{Linearized Boltzmann Equation and Moment Equation}

We derive a moment equation from a linearized Boltzmann equation for a two-component mass and population imbalanced normal Fermi gas by extending formulation for a mass and population balanced normal Fermi case~\cite{Watabe2009,Watabe2010}. 
Let $\sigma=\uparrow$ or $\downarrow$ be a pseudo-spin describing each component of the Fermi gas with an atomic mass $m_\sigma$. The Boltzmann equation for a distribution function $f_\sigma ({\bf p}, {\bf r}, t)$ is given by 
 \begin{align}
 \frac{\partial f_{\sigma}({\bf p},{\bf r},t)}{\partial t}
  +\frac{ 1 }{m_{\sigma}} {\bf p} \cdot \nabla_{{\bf r}} f_{\sigma}({\bf p},{\bf r},t) & 
  \nonumber 
  \\ 
  -\nabla_{{\bf r}} U_{\sigma}({\bf r},t) \cdot \nabla_{{\bf p}} f_{\sigma}({\bf p},{\bf r},t)
  & = {\mathcal I} [f_{\sigma}], 
  \label{eq:1}
 \end{align}
 where $\mathcal{I} [f_{\sigma}]$ is a collision integral term, and $U_{\sigma}({\bf r},t) \equiv g n_{-\sigma}({\bf r},t) +U_{\text{ext},\sigma}({\bf r},t)$ is the sum of the mean-field interaction potential with a coupling strength $g$ and an external perturbation potential $U_{\text{ext},\sigma}$. Here, $n_{\sigma}({\bf r},t) =  \displaystyle \int \cfrac{d{\bf p}}{(2\pi\hbar)^{3}} f_{\sigma}({\bf p},{\bf r},t)$ is a local density of the pseudo-spin $\sigma$-component. 

In order to investigate the small amplitude oscillations around the static equilibrium, 
it is convenient to write the distribution function as
$f_{\sigma}({\bf p},{\bf r},t) \equiv f_{\sigma}^0({\bf p}) + \delta f_{\sigma}({\bf p},{\bf r},t) \equiv  f_{\sigma}^0({\bf p}) + \cfrac{\partial f^0_{\sigma}}{\partial \epsilon^0_{\sigma}} \nu_{\sigma}({\bf p},{\bf r},t)$,
where $f_{\sigma}^0({\bf p})$ is the static equilibrium value.
Using the above expression in the Boltzmann equation (\ref{eq:1}) and linearizing it in $\nu_{\sigma}$,
we obtain the linearized Boltzmann equation
 \begin{align}
 &  \frac{\partial f^0_{\sigma}}{\partial \epsilon^0_{\sigma}}
  \biggl \{ \frac{\partial \nu_{\sigma}({\bf p},{\bf r},t)}{\partial t}
  +\frac{ 1 }{m_{\sigma}} {\bf p} \cdot \nabla \nu_{\sigma}({\bf p},{\bf r},t) 
  \nonumber 
  \\
 &  -\frac{ 1 } {m_{\sigma}} {\bf p} \cdot \nabla \biggl[ g \delta n_{-\sigma}({\bf r},t)  +U_{\text{ext},\sigma}({\bf r},t) \biggr]
  \biggr \} 
  \nonumber   \\ 
  = & -\frac{1}{\tau}\frac{\partial f^0_{\sigma}}{\partial \epsilon^0_{\sigma}} [ \nu_{\sigma} ({\bf p},{\bf r},t) - \tilde \nu_{\sigma} ({\bf p},{\bf r},t) ] , 
    \label{eq:7}
 \end{align}
where we have introduced the density fluctuation $\delta n_{\sigma} ({\bf r},t) \equiv \displaystyle \int \cfrac{d{\bf p}}{(2\pi\hbar)^{3}} \delta f_{\sigma}({\bf p},{\bf r},t)$.
In the right hand side of Eq.~(\ref{eq:7}), we have adopted a relaxation time approximation, 
where we employ the mean-collision time as the relaxation time $\tau $ that characterizes the time when the gas relaxes to the local equilibrium $\tilde f_{\sigma}$, the detail of which is summarized in Appendix ~\ref{AppendixA}. 
Here, $\cfrac{\partial f^0_{\sigma}}{\partial \epsilon^0_{\sigma}} \tilde \nu_{\sigma}({\bf p},{\bf r},t) \equiv \tilde f_{\sigma} ({\bf p}, {\bf r}, t) - f_{\sigma}^{0} ({\bf p}, {\bf r}, t) \equiv \delta \tilde f_{\sigma}({\bf p}, {\bf r}, t)$ is the linearized form of the local equilibrium distribution function.
The local equilibrium distribution function that satisfies ${\cal I}[\tilde f_{\sigma}]=0$ is given by 
 \begin{align}
  \tilde{f}_{\sigma}({\bf p})
  =\cfrac{1}{\exp \left [ \tilde{\beta}_{\sigma}  ({\bf p} - m_{\sigma} \tilde{{\bf v}}_{\sigma})^2 /( 2m_{\sigma} ) \right ]  \tilde z^{-1}_{\sigma} +1 }, 
  \label{eq:3}
  \end{align}
where the local temperature ${1} / {\tilde{\beta}_{\sigma} } \equiv k_B \tilde{T}_{\sigma}  \equiv \tilde{\theta}_{\sigma} $, the local velocity $\tilde{{\bf v}}_{\sigma} $, and the local fugacity $\tilde z_{\sigma}  \equiv \exp [ \tilde{\beta}_{\sigma} (\tilde{\mu}_{\sigma}   - g  \tilde{n}_{-\sigma} ) ]$ with the local density $\tilde{n}_{\sigma} $ and the local chemical potential $\tilde \mu_{\sigma} $ all 
depend on the position and time.
Expanding Eq.~(\ref{eq:3}) with respect to small fluctuations $\delta \tilde{\theta}  \equiv \tilde{\theta} - \theta^0 $, $ \delta \tilde{\mu}_{\sigma} \equiv \tilde{\mu}_{\sigma} - \mu^0_{\sigma}$, as well as $\delta \tilde{{\bf v}} = \tilde{{\bf v}}  - {\bf v}^0 =  \tilde{{\bf v}} $ around those static equilibrium quantities $\theta^0=k_BT^0$, $\mu^0_{\sigma}$, and ${\bf v}_{\sigma}^0=0$, 
we find the linearized form of the local equilibrium distribution function as
 \begin{align}
\delta \tilde f_{\sigma}({\bf p}, {\bf r}, t) = & 
\frac{\partial f^0_{\sigma}}{\partial \epsilon^0_{\sigma}} [ \tilde a_{\sigma}({\bf r},t)+ \tilde{\bf b}_{\sigma}({\bf r},t) \cdot {\bf p} + \tilde c_{\sigma}({\bf r},t)p^2], 
\label{eq:dtf}
\end{align} 
where the parameters $\tilde a_{\sigma},\tilde{\bf b}_{\sigma},\tilde c_{\sigma}$ are related to the local quantities through
(notify that a tilde is used for a local equilibrium quantity in this paper, and here after, in a quantity with the tilde, we will not explicitly write the $({\bf r}, t)$-dependence as well as the $({\bf q}, \omega)$-dependence in their Fourier-space)  
 \begin{align}
  \tilde a_{\sigma} 
  \equiv&
  -\beta^0_{\sigma} (g \ n^0_{-\sigma} -\mu^0_{\sigma}) \delta \tilde{\theta}_{\sigma} 
  -\delta \tilde{\mu}_{\sigma} 
  +g  \delta \tilde{n}_{-\sigma} ,
  \label{eq5} 
  \\
  \tilde {\bf b}_{\sigma} 
  \equiv& - \delta \tilde{{\bf v}}_{\sigma} ,
    \label{eq6} 
    \\
   \tilde c_{\sigma} 
  \equiv& -\frac{\beta^0_{\sigma}}{2m_{\sigma}}\delta \tilde{\theta}_{\sigma}. 
    \label{eq7} 
  \end{align} 
We consider an external potential that excites modes of frequency $\omega$ and wavevector ${\bf q}$, namely
$U_{{\rm ext},\sigma}({\bf r},t)=U_{{\rm ext},\sigma}({\bf q},\omega)e^{i({\bf q}\cdot {\bf r}-\omega t)}$.
In this case, the plane-wave solution of Eq.~(\ref{eq7}) given by $\nu_{\sigma}({\bf p},{\bf r},t) =\nu_{\sigma}({\bf p},{\bf q},\omega) \exp{ [  i({\bf q} \cdot {\bf r} - \omega t) ] }$ gives
 \begin{align}
 \frac{\partial f^0_{\sigma}}{\partial \epsilon^0_{\sigma}} &\Biggl\{
 \Bigl(\omega-\frac{{\bf p} \cdot {\bf q}}{m_{\sigma}} \Bigr)\nu_{\sigma}({\bf p})
  +\frac{{\bf p} \cdot {\bf q}}{m_{\sigma}}
  \Bigl[ 
 g \delta n_{-\sigma} 
  +U_{\text{ext},\sigma} \Bigr]
  \Biggr\}
  \nonumber 
  \\
 & =-\frac{i}{\tau} \frac{\partial f^0_{\sigma}}{\partial \epsilon^0_{\sigma}} \biggl[\nu_{\sigma}
  -\Bigl( \tilde a_{\sigma}+ \tilde {\bf b}_{\sigma} \cdot {\bf p} +  \tilde c_{\sigma}p^2
  \Bigr) \biggr]. 
  \label{eq:20}
 \end{align}

 This linearized Boltzmann equation can be solved by introducing spherical harmonics 
 \begin{align}
  \nu_{\sigma}({\bf p}) \equiv \sum^{\infty}_{l=0} \sum_{m=-l}^{l}
  \nu^m_{\sigma ,l}(p) P^m_l(\cos \theta)e^{im \phi}.  
 \end{align} 
Since we are considering an isotropic ($s$-wave) interaction and ordinary density/spin modes, we take the mode $m = 0$, and omit the index $m =0$ hereafter. 
By multiplying $p^nP_{l'}^{}(\cos \theta) / (2\pi\hbar)^3$ and integrating over ${\bf p}$, we obtain the moment equation 
  \begin{align}
  &  \omega \langle p^n \nu_{\sigma,l} \rangle
   -\frac{l}{2l-1} \frac{q}{m_{\sigma}} \langle p^{n+1} \nu_{\sigma,l-1} \rangle
   -\frac{l+1}{2l+3} \frac{q}{m_{\sigma}} \langle p^{n+1} \nu_{\sigma,l+1} \rangle
   \nonumber\\
   & +\frac{q}{m_{\sigma}} W_{\sigma,n+1} \delta_{l,1}
   \bigl(
 g \langle \nu_{-\sigma,0} \rangle
   +U_{\text{ext},\sigma}
   \bigr)
   \nonumber\\
   =&-\frac{i}{\tau} 
   \biggl [ 
   \langle p^{n} \nu_{\sigma,l} \rangle
-    ( C_{N,\sigma} \langle \nu_{\sigma,0} \rangle +    C_{N,-\sigma} \langle \nu_{-\sigma,0} \rangle ) \delta_{l,0}
   \nonumber\\
   &- C_{P,\sigma}
(    \langle p\nu_{\uparrow,1} \rangle  +    \langle p\nu_{\downarrow,1} \rangle ) 
   \delta_{l,1}
   \nonumber\\   
   &-  C_{E, \sigma}
   \biggl(
   \frac{\langle p^2\nu_{\uparrow,0} \rangle}{ m_{\uparrow}}
   +\frac{\langle p^2\nu_{\downarrow,0} \rangle}{ m_{\downarrow}}
   \biggr)
   \delta_{l,0}
      \biggr ], 
   \label{eq:moment-eq}
  \end{align}
where we have defined 
  \begin{align}
   \langle p^n\nu_{\sigma,l} \rangle
   \equiv & 
   \int \frac{d{\bf p}}{(2\pi \hbar)^3}
   \frac{\partial f^0_{\sigma}}{\partial \epsilon^0_{\sigma}} p^n \nu_{\sigma,l}(p), 
   \\ 
    W_{\sigma,n} \equiv & \int \frac{d {\bf p}}{(2\pi \hbar)^3} \frac{\partial f^0_{\sigma}}{\partial \epsilon ^0_{\sigma}} p^n , 
    \\ 
    C_{N,\sigma} \equiv &  
   \frac{W_{\sigma,n}}{W_{\sigma,0}}
   +\frac{1}{m_{\sigma}^2   \Phi}
   \biggl(
   \frac{W_{\sigma,n}W_{\sigma,2}^2}{W_{\sigma,0}^2}
   -\frac{W_{\sigma,n+2}W_{\sigma,2}}{W_{\sigma,0}}
   \biggr) , 
   \\ 
   C_{N,-\sigma} \equiv & 
   \frac{1}{m_{\sigma} m_{-\sigma}  \Phi}
   \biggl(
   \frac{W_{\sigma,n}W_{\sigma,2}W_{-\sigma,2}}{W_{\sigma,0}W_{-\sigma,0}}
   -\frac{W_{\sigma,n+2}W_{-\sigma,2}}{W_{-\sigma,0}}
   \biggr) , 
   \\ 
   C_{P,\sigma} \equiv &    \frac{W_{\sigma,n+1}}{(\sum_{\sigma} W_{\sigma,2})} , 
   \\ 
   C_{E, \sigma} \equiv & 
     \frac{1}{m_{\sigma} \Phi}
   \biggl(
   W_{\sigma,n+2}
   - \frac{W_{\sigma,2}W_{\sigma,n}}{W_{\sigma,0}}
   \biggr) , 
   \\ 
     	\Phi
 	\equiv & 
	\sum \frac{1}{ m_{\sigma}^2} \Biggl(
 	W_{\sigma,4} -\frac{W_{\sigma,2}^2}{W_{\sigma,0}}
 	\Biggr) . 
  \end{align} 
Note that the density fluctuation is given by $\delta n_{\sigma} = \langle \nu_{\sigma, 0} \rangle$, and the coefficients $\tilde a_{\sigma}$, $\tilde b_{\sigma} \equiv | \tilde {\bf b}_{\sigma}|$, and $\tilde c_{\sigma}$ have been determined by the conservation law of number of particles, momentum, as well as energy, the condition of which is summarized in Appendix~\ref{AppendixB}. These conservation laws are required for the collision integrals. As a consequence, the other moments, such as the spin current, are not conserved by the collision.

If we solve (\ref{eq:moment-eq}) with $U_{{\rm ext}, \sigma} = 0$, the moment equation  (\ref{eq:moment-eq})  can be regarded as an eigenvalue equation, which provides eigenmode frequencies $\omega$ as well as eigenvectors $\langle p^{n} \nu_{\sigma, l} \rangle$. The eigenfrequencies in the hydrodynamic and collisionless limits are discussed in Appendices~\ref{AppendixC} and~\ref{AppendixD}, respectively. 
If we solve (\ref{eq:moment-eq}) with $U_{{\rm ext}, \sigma} \neq 0$, the moment equation  (\ref{eq:moment-eq}) can be regarded as simultaneous linear equations, which provides $\langle p^{n} \nu_{\sigma, l} \rangle$ as a function of $\omega$. 

The density (in-phase) mode is characterized by the sum of each density fluctuation, $\delta n = \delta n_{\uparrow} + \delta n_{\downarrow}$, which can be simply excited by an in-phase external perturbation $(U_{{\rm ext},\uparrow}, U_{{\rm ext},\downarrow}) = (1, 1) U_{{\rm ext}}^{\rm d}$. 
Since the density fluctuation in the $\sigma$-component is given by the zeroth moment $\delta n_{\sigma}  = \langle \nu_{\sigma, 0}  \rangle$, 
the density response function $\chi_{\rm d} ({\bf q}, \omega)$ is then obtained from the zeroth moment, given by 
\begin{align}
\chi_{\rm d} ({\bf q}, \omega) \equiv \frac{ \langle \nu_{\uparrow, 0} ({\bf q}, \omega)  \rangle + \langle \nu_{\downarrow, 0} ({\bf q}, \omega)  \rangle   }{ U_{\rm ext}^{\rm d} ({\bf q}, \omega) }. 
\end{align}
On the other hand, 
the spin (out-of-phase) mode is characterized by the difference of each density fluctuation, $\delta s = \delta n_{\uparrow} - \delta n_{\downarrow}$, which can be simply excited by an out-of-phase external perturbation $(U_{{\rm ext},\uparrow}, U_{{\rm ext},\downarrow}) = (1, - 1) U_{{\rm ext}}^{\rm s}$. 
The spin response function $\chi_{\rm s} ({\bf q}, \omega)$ is now related to the moment as 
\begin{align}
\chi_{\rm s} ({\bf q}, \omega) \equiv \frac{ \langle \nu_{\uparrow, 0} ({\bf q}, \omega)  \rangle - \langle \nu_{\downarrow, 0} ({\bf q}, \omega)  \rangle }{ U_{\rm ext}^{\rm s} ({\bf q}, \omega) }. 
\end{align}
Using these response functions, we calculate the density and spin dynamic structure factor 
\begin{align}
S_{\rm d,s} ({\bf q}, \omega) = - \frac{1}{\pi} \frac{1}{1 - \exp (- \beta \hbar \omega)} {\rm Im} \chi_{\rm d, s} ({\bf q}, \omega) . 
\end{align}

\section{results}

\begin{figure}[tbp] 
\begin{center}
\includegraphics[width=8cm]{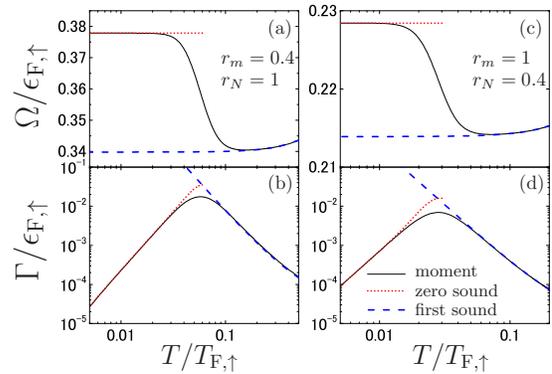}
\end{center} 
\caption{
Temperature dependence of the frequency $\Omega$ and the damping rate $\Gamma$ in the density collective mode. 
The frequency $\Omega$ (a) and the damping rate $\Gamma$ (b) in a mass imbalanced gas. 
The frequency $\Omega$ (c) and the damping rate $\Gamma$ (d) in a population imbalanced gas. 
We have used $g = 15 \epsilon_{{\rm F},\uparrow}V/N_{\uparrow}$, and $q = 0.05 k_{{\rm F},\uparrow}$, where $k_{{\rm F}, \uparrow}$ is the Fermi wavenumber of the majority component. 
 }
\label{fig1}
\end{figure}

An eigenvalue obtained from our moment method (\ref{eq:moment-eq}) clearly shows the crossover from the collisionless zero sound mode to the hydrodynamic first sound mode both in the mass imbalanced case $r_{m} \equiv m_{\downarrow} / m_{\uparrow} \neq 1$ and in the population imbalance case $r_{N} \equiv N_{\downarrow} / N_{\uparrow} \neq 1$ (Fig.~\ref{fig1}). In this paper, we always take the pseudo-spin $\uparrow$-component as the majority component of the population or the heavier mass component, which provides $0 < r_{m, N} \leq 1$. The normalization is also performed by using quantities in the majority component, such as the Fermi energy $E_{{\rm F},\uparrow}$ and the Fermi temperature $T_{{\rm F}, \uparrow}$. In the low (high) temperature regime, frequency and damping rate of a eigenvalue excellently reproduce those obtained from equations in the collisionless (hydrodynamic) regime (\ref{eq:z1}) and (\ref{eq:z2}) ((\ref{eq:fs-1}), (\ref{eq:fs-2}) and (\ref{eq:fs-3})). 

\begin{figure}[tbp] 
\begin{center}
\includegraphics[width=8cm]{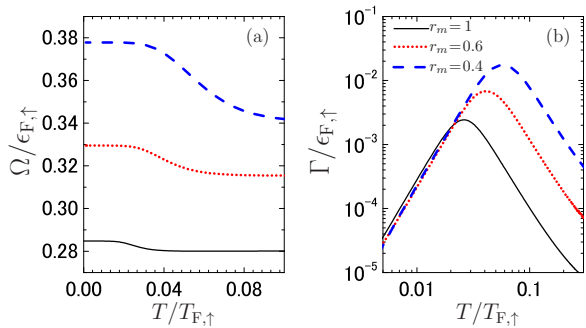}
\end{center} 
\caption{
Mass imbalance dependence of the frequency $\Omega$ (a) and the damping rate $\Gamma$ (b) in the density collective mode, with $g = 15 \epsilon_{{\rm F},\uparrow}V/N_{\uparrow}$ and $q = 0.05 k_{{\rm F},\uparrow}$. 
 }
\label{fig2}
\end{figure}

\begin{figure}[tbp] 
\begin{center}
\includegraphics[width=8cm]{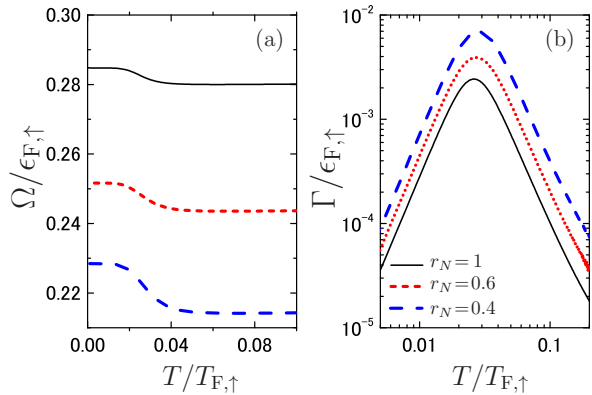}
\end{center} 
\caption{
Population imbalance dependence of the frequency $\Omega$ (a) and the damping rate $\Gamma$ (b) in the density collective mode, with $g = 15 \epsilon_{{\rm F},\uparrow}V/N_{\uparrow}$ and $q = 0.05 k_{{\rm F},\uparrow}$. 
 }
\label{fig3}
\end{figure}

Difference between mass imbalance effect and population imbalance effect is clearly seen if we change imbalance parameters $r_{m,N}$ (Figs.~\ref{fig2} and~\ref{fig3}). 
As mass imbalance effect is stronger with $r_{m}$ decreasing, the sound speed becomes faster in all the temperature regimes, which may be consistent with the form (\ref{fsomega2}), and the damping rate grows in the crossover and the hydrodynamic regime. 
The crossover region is shifted to higher temperatures due to the mass imbalance effect. 
Since the zero sound collective mode can appear in the very low temperature regime, which is hard to reach, the mass imbalance effect may be one of useful routes to study the collisionless zero sound mode. 

\begin{figure}[tbp] 
\begin{center}
\includegraphics[width=7cm]{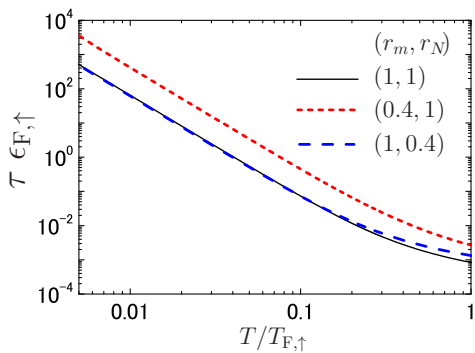}
\end{center} 
\caption{
Mass and population imbalance dependence of the mean-collision time with $g = 15 \epsilon_{{\rm F},\uparrow}V/N_{\uparrow}$. 
 }
\label{fig4}
\end{figure}

On the other hand, 
as population imbalance effect is stronger with $r_{N}$ decreasing, the sound speed becomes slower in all the temperature regimes, which may be consistent with the form (\ref{fsomega2}) with the weight $W_{\downarrow,n}$ possibly decreasing, and the damping rate grows also in all temperature regimes. 
The crossover region is not shifted to higher temperatures by the population imbalance effect, which is stark contract with the mass imbalance case. 
This contrast is caused by the difference between imbalance effects on the relaxation time (Fig.~\ref{fig4}). 
The mass, rather than population, imbalance effect drastically enhances the mean-collision time, although the population imbalance effect also quantitatively modifies the mean-collision time. This mass imbalance effect provides the shift of the crossover regime to higher temperatures.

\begin{widetext}
\begin{figure}[tbp] 
\begin{center}
\includegraphics[width=18cm]{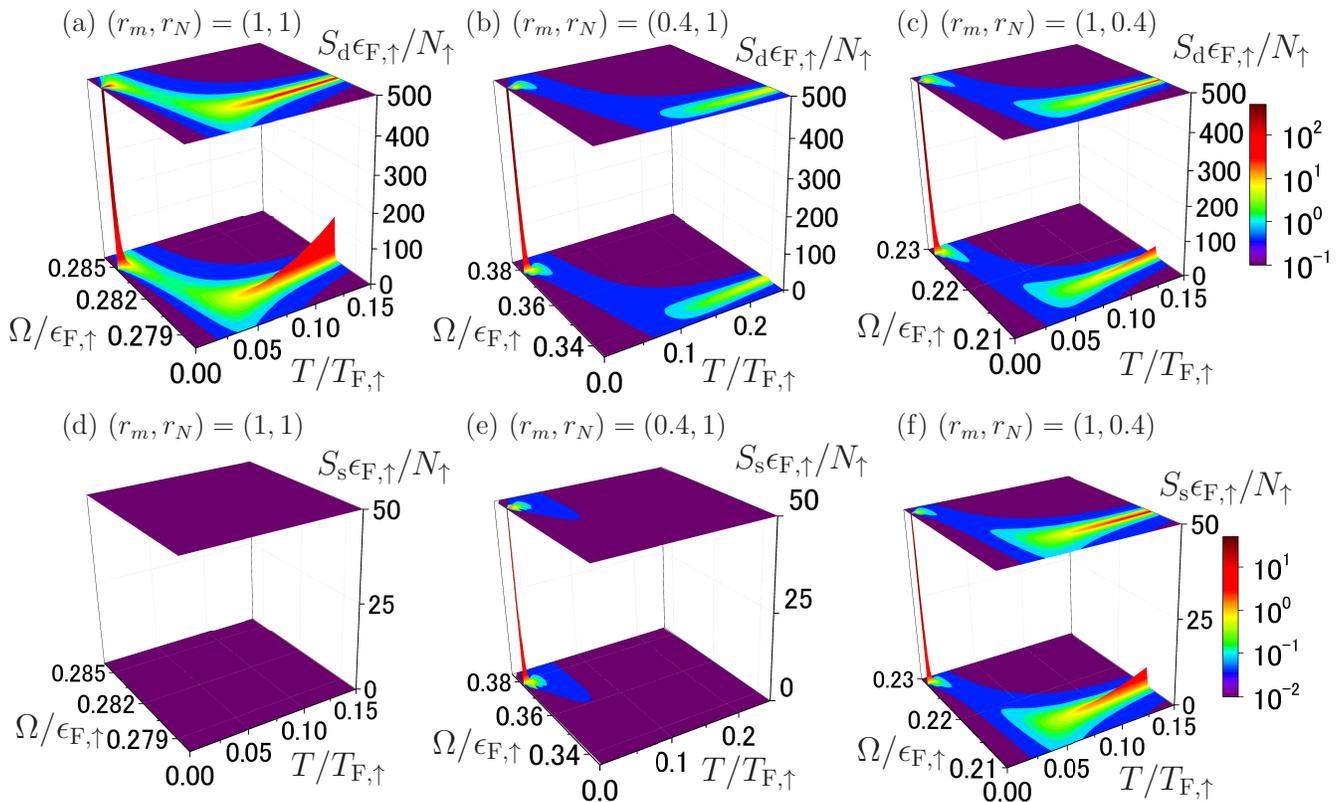}
\end{center} 
\caption{
Temperature and frequency dependence of the density dynamic structure factor in a population balanced gas (a), a mass imbalanced gas (b) and a population imbalanced gas (c). 
The spin dynamic structure factor in a population balanced gas (d), a mass imbalanced gas (e) and a population imbalanced gas (f). 
We have used $g = 15 \epsilon_{{\rm F},\uparrow}V/N_{\uparrow}$, and $q = 0.05 k_{{\rm F},\uparrow}$. 
 }
\label{fig5}
\end{figure}
\end{widetext}

The mass and population imbalance effect is also clearly appears in the spin dynamic structure factor, rather than the density dynamic structure factor (Fig.~\ref{fig5}). 
In the density dynamic structure factor $S_{\rm d}$, quantitative difference alone appears among a balanced gas, a mass imbalanced gas and a population imbalance gas. 
In the density dynamic structure factor $S_{\rm d}$, pronounced peaks emerge in the collisionless regime as well as in the hydrodynamic regime, the positions of which are quantitatively changed by imbalance effect as shown in Figs.~\ref{fig2} and~\ref{fig3}. 
In the hydrodynamic regime, the peak strength in mass and population imbalanced gases becomes weaker than that in the balanced case. 
In the crossover regime, strength of the peak becomes weak and wide because of the large damping rate. 

In the spin dynamic structure factor $S_{\rm s}$, qualitative difference can be clearly seen among a balanced gas, a mass imbalanced gas as well as a population imbalanced gas. 
Since the relaxation to the local equilibrium is strongly suppressed in the low temperature collisionless regime, the out-of-phase spin mode is expected to occur. 
Indeed, although in a balanced gas any structures of $S_{\rm s}$ can not be seen because density fluctuations in each pseudo-spin component are exactly canceled out each other, those can be seen both in mass and population imbalanced gases. 
In the hydrodynamic regime, the qualitative difference can be seen between the mass and population imbalanced gasses.  
%
%
%
%
In a mass imbalanced but population balanced gas, the spin dynamic structure factor does not hold any peak in the hydrodynamic regime. In this hydrodynamic limit, the main contribution for a collective mode is given by thermodynamic quantities: number of particles for each spin component, the total energy, and the velocity. Since the number of particles for each component is balanced, the spin response is suppressed in a mass imbalanced but population balanced gas as shown in Fig.~\ref{fig5}. We note that, because of the non-conservation of the spin current by the collision, the distinct peak for the spin diffusion mode emerges in the spin dynamic structure factor at $\omega =0$. 
In a mass balanced but population imbalanced gas, the spin dynamic structure factor can hold a peak structure in the hydrodynamic regime. Since each component has different strength of density fluctuations in a population imbalanced gas, spin dynamic structure factor may revival in the hydrodynamic regime. If we assume the case where the population imbalance is very large, one may expect that the density and spin response functions behave similarly, the tendency of which is clearly seen in panels (c) and (f) in Fig.~\ref{fig5}. 

\onecolumngrid

\twocolumngrid

\begin{figure}[tbp] 
\begin{center}
\includegraphics[width=8cm]{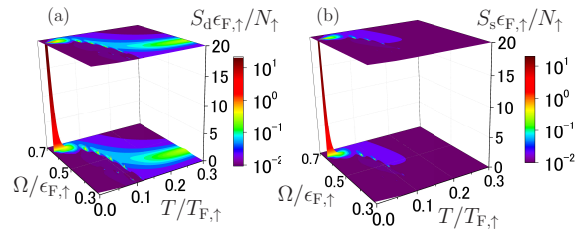}
\end{center} 
\caption{
Temperature and frequency dependence of the density dynamic structure factor (a) and the spin dynamic structure factor (b) in a mass imbalanced $^{6}$Li and $^{40}$K fermionic mixture. 
We have used $g = 5 \epsilon_{{\rm F},\uparrow}V/N_{\uparrow}$, and $q = 0.05 k_{{\rm F},\uparrow}$. 
 }
\label{fig6}
\end{figure}

In Figs. \ref{fig1}-\ref{fig5}, we have chosen a large coupling constant $g = 15 \epsilon_{{\rm F},\uparrow}V/N_{\uparrow}$, 
for the purpose of illustration of the mass and population imbalance effects on the eigenvalue in the moment equation. 
In the large coupling constant case, we can easily extract the eigenvalue of the collective mode, 
since it is well separated from the eigenvalues of the single particle excitations forming continuum~\cite{Watabe2010}. In the small coupling constant case, the eigenvalue of the zero sound becomes close to the continuum, and the eigenvalue of the collective mode in the crossover regime is positioned in the continuum~\cite{Watabe2009}. 
However, if we calculate the dynamic structure factor, the structure of the collective mode can be clearly seen. 
In Fig.~\ref{fig6}, we calculated the temperature and frequency dependence of the density dynamic structure factor and the spin dynamic structure factor in a mass imbalanced $^{6}$Li and $^{40}$K fermionic mixture in the smaller coupling constant case. 
The structure of the density and spin dynamic structure factors with the smaller coupling constant (Fig.~\ref{fig6}) are qualitatively the same as those with the large coupling constant in (b) and (e) of Fig.~\ref{fig5}. In this sense, our results using the large coupling constant are useful for a qualitative illustration of physics, even for a weaker interaction case. 
We note that the weaker coupling case is also studied for a balanced Fermi gas in Ref.~\cite{Watabe2010}, and the significant sharp peak, which reflects the zero sound, emerges slightly above the continuum in the low temperature region. 

Recently, a homogeneous atomic Fermi gases are realized in a uniform potential~\cite{Mukherjee2017}. 
If red and blue detuned localized potential is selectively applied to each component, similarly to the experiment for the density sound propagation~\cite{Kinast2004,Bartenstein2004,Joseph2007}, in-phase and out-of-phase fluctuation may be excited, and rich imbalance effect might be observed. 
Mass and population imbalance effect on spin sound mode unveiled in this paper is an interesting and challenging problem in an imbalanced two-component normal Fermi gas. 
We note that it is also an interesting issue to unveiled the population and mass imbalanced effect on the collective modes of the trapped Fermi gases, such as the breathing mode, and the quadrupole mode. 

We summarize other interesting subjects for future studies. 
We have used the simplified contact interaction for the large coupling constant. 
In order to study the strong coupling case quantitatively, we should extend our moment method to include the momentum-dependent cross section used in Ref.~\cite{Chiacchiera2009}, which is valid for the unitarity regime. 
Other prospective studies include the more generalized extension of the moment method to include the higher Landau parameters and the spin Landau parameters. In the repulsive contact interaction used in this study, we can find that the spin Landau parameter is negative, which can be easily checked by the Hartree-Fock approximation~\cite{Ueda2010}. It provides the absence of the spin zero sound in the balanced gas. 
We also note that, because of the $s$-wave contact interaction, the transverse zero sound is also absent in our calculation. 
Exploring the above mentioned possibilities for variant Fermi gasses is an interesting issue. 

In the mean-field theory for the large repulsive coupling constant case, the Stoner instability for the ferromagnetism may be considered~\cite{Jo2009}, which can be discussed in the random phase approximation at $T=0$. As discussed in Ref.~\cite{Watabe2009}, our moment method in the collisionless limit can reproduce the results given by the random phase approximation in the long-wavelength regime. In this sense, it may be possible to discuss the so-called Stoner instability in the strong repulsive interaction case in the moment method. On the other hand, we focus on the nonzero temperature case and the beyond mean-field effect is included through the relaxation time approximation in our theory. The response function with this approximation at nonzero temperature always has the imaginary part and does not diverge unlike the case of the Stoner instability in our calculation. Within our theoretical framework, we do not found the signature of the instability, but found that of the collective mode in the density and spin response function. 
In ultracold atomic gases, the basic Stoner model cannot be realized, because of the decay into bound pairs in a lower branch~\cite{Zhang2011,Sanner2012}, where the order of the decay rate is $0.1 E_{\rm F}/\hbar$~\cite{Sanner2012}. Although it prevents the study of equilibrium phases and hydrodynamics of Fermi gases with very large repulsive interaction, it may provide the possibility to study the collisionless zero sound in the strongly repulsive Fermi gas, if we control the collective mode frequency to be relatively large compared with the decay rate into bound pairs, as shown in Fig.~\ref{fig6} for the zero sound collisionless regime. The study for finding the parameters for realizing the collective mode in the experiment with both weak and strong repulsive interaction cases is a remaining important issue. 

The other prospective studies are to understand the solution of the moment method more clearly. 
Our method is constructed to satisfy the conservation laws of the number of particles, momentum and energy. 
However, the relation to the sum-rule for the dynamic structure factor is not clear. Furthermore, it is an open and complicated problem to understand meaning of all the eigenmodes only from the structure of the eigenvectors. This is the reason we focus on the eigenvalues and the response functions obtained from $\delta n_{\uparrow} \pm \delta n_{\downarrow}$. To understand the structure of the eigenvectors in the moment method provides deeper understanding of the excitations given by the Boltzmann equation. 

\section{Conclusions}

      We have studied mass and population imbalance effect on density (in-phase) and spin (out-of-phase) collective modes in a two-component normal Fermi gas. We derived a moment equation for a mass and population imbalanced Fermi gas from the linearized Boltzmann equation. This moment equation can provide not only frequency and damping rate of the collective mode, but also the density/spin dynamic structure factor. We have shown that imbalance effect offer rich crossover behavior from collisionless to hydrodynamic regimes. 
The mass imbalance effect increases a collective mode frequency, and shifts the crossover regime to higher temperature regimes because of enhancement of the mean-collision time. 
On the other hand, the population imbalance effect decreases a collective mode frequency and enhances the damping rate; however, the crossover regime does not drastically change as in the case in the mass imbalanced gas. 
The spin dynamic structure factor shows a variety of behaviors: for a mass and population imbalanced gas, a pronounced peak emerges in the collisionless regime, which is not seen in a balanced gas at all temperatures. 
Since the strength of the density fluctuation is different between two-components in a population imbalanced gas, a weak structure emerges in the hydrodynamic regime in the population imbalanced gas, which is contrast with the mass imbalanced gas. 

\begin{acknowledgments}
S.W. was supported by JSPS KAKENHI Grant No. JP16K17774.
T.N. was supported by JSPS KAKENHI Grant No. JP16K05504.
\end{acknowledgments}

\appendix

\section{relaxation time}\label{AppendixA}

The mean-collision time $\tau$, which is employed as the relaxation time in this paper, is given by 
\begin{widetext}
\begin{align}
 \frac{N_{\text{tot}}}{V}
 \frac{1}{\tau}
 \equiv&
 \frac{2 \pi g^2}{\hbar}
 \int \frac{d{\bf p}_1}{(2 \pi \hbar)^3}
 \int \frac{d{\bf p}_2}{(2 \pi \hbar)^3}
 \int \frac{d{\bf p}_3}{(2 \pi \hbar)^3}
 \int d{\bf p}_4 \delta({\bf p}_1 + {\bf p}_2 -{\bf p}_3 -{\bf p}_4)
 \delta
 \biggl(
 \frac{p_1^2}{2m_{\uparrow}}
 +\frac{p_2^2}{2m_{\downarrow}}
 -\frac{p_3^2}{2m_{\downarrow}}
 -\frac{p_4^2}{2m_{\uparrow}}
 \biggr)
 \nonumber\\
 &\hspace{6cm}
 \times
 [1 - \tilde{f}_{\uparrow}(1)]
 [1 - \tilde{f}_{\downarrow}(2)]
 \tilde{f}_{\downarrow}(3) \tilde{f}_{\uparrow}(4), 
\end{align} 
\end{widetext}
where we used a convention $f_{\sigma}(i) \equiv f_{\sigma}({\bf p}_i,{\bf r},t)$. 
From this collision integral term  $\mathcal{I} [\tilde{f}_{\sigma}]$, one can find that the spin current is not conserved. 
Once we introduce the total mass $M \equiv m_{\uparrow} + m_{\downarrow}$, the reduced mass $m_{\rm re} \equiv m_{\uparrow}m_{\downarrow}/M$, momenta of the center of mass ${\bf P}$ and ${\bf P}'$, as well as those of the relative motion ${\bf p}$ and ${\bf p}'$, 
the mean-collision time is reduced into 
 \begin{align}
 \frac{N_{\text{tot}}}{V}
 \frac{1}{\tau}
  =&  \frac{m_{\rm re} g^2}{(2\pi)^{5}\hbar^{10}}
  \int_{0}^{\infty} P^2  dP
  \int_{0}^{\pi} d\theta_P  \sin \theta_P 
  \\ 
  & \times 
  \int p^3 dp
  \int_{-1}^{1} dy 
  \int_{-1}^{1} dy'  
  F. 
  \end{align}
Here, we have used relations 
\begin{align}
 {\bf p}_1 =  \frac{m_{\uparrow}}{M} \boldsymbol{P} +   \boldsymbol{p}, & \quad 
 {\bf p}_2 =  \frac{m_{\downarrow}}{M} \boldsymbol{P}  -  \boldsymbol{p}, 
 \\ 
 {\bf p}_3 =  \frac{m_{\downarrow}}{M} \boldsymbol{P}'  -  \boldsymbol{p}', & \quad 
 {\bf p}_4 =  \frac{m_{\uparrow}}{M} \boldsymbol{P}' +  \boldsymbol{p}', 
 \label{eq:a150}
\end{align}
the momentum conservation law of center of mass ${\bf P} = {\bf P}'$, as well as the energy conservation law of the relative motion ${\bf p}^{2}/(2m_{\rm re}) = {\bf p}'^{2}/(2m_{\rm re})$. 
We have also introduced the polar coordinate in the momentum space, and defined
$y$ and $y'$ by ${\bf P} \cdot {\bf p} \equiv P p y$ as well as ${\bf P}' \cdot {\bf p}' \equiv P' p' y'$. The function $F$ is then given by 
\begin{align}
F 
 \equiv &\frac{1}{4}
 \frac{1}{
 \cosh (A) +  \cosh (By + C)
 }
  \frac{1}{
 \cosh(A) +  \cosh (By' + C)
  }, 
\end{align}
 where 
 \begin{align}
 A \equiv &
 \frac{\tilde{\beta}}{2} \biggl(
 \frac{\boldsymbol{P}^2}{2M} +\frac{\boldsymbol{p}^2}{2 m_{\rm re}} -  \tilde{\mu}_{\uparrow}^{\text{conv}} - \tilde{\mu}_{\downarrow}^{\text{conv}}
 \biggr), 
 \\
 B \equiv &
 \frac{\tilde{\beta}}{M} P p, 
 \\
 C \equiv &
 \frac{\tilde{\beta}}{2}
 \biggl(
 \frac{m_{\uparrow} -m_{\downarrow}}{M} \frac{\boldsymbol{P}^2}{2M}
 -\frac{m_{\uparrow} -m_{\downarrow}}{M}  \frac{\boldsymbol{p}^2}{2 m_{\rm re}}
 - \tilde{\mu}_{\uparrow}^{\text{conv}} + \tilde{\mu}_{\downarrow}^{\text{conv}} 
 \biggr) , 
\end{align}
with $\tilde{\mu}_{\sigma}^{\text{conv}} \equiv \tilde{\mu}_{\sigma} - g \tilde{n}_{-\sigma}$. 
The integral with respect to $y$ is analytically preformed as follows:
\begin{align}
  & \frac{1}{2}
  \int_{-1}^{1} \frac{dy}{\cosh(A)+\cosh(By +C)}
   \\ 
  =&
    \frac{1}{B \sinh{A}}
  \ln
  \biggl [ 
  \frac{\cosh{(A +B)} +\cosh{C}}{\cosh{(A -B)} +\cosh{C} }
  \biggr ], 
 \end{align}
which is also the case for the integral with respect to $y'$. 
As a result, the relaxation time, defined by the mean-collision time, reads 
 \begin{align}
 \frac{N_{\text{tot}}}{V}
 \frac{1}{\tau} 
 = & 
 \frac{g^2  m_{\rm re} M^2}{16 \pi^5 \hbar^{10}}
 (k_B \tilde{T})^2
 \int_0^{\infty}  d P
 \int_0^{\infty} d p p 
 \nonumber 
 \\ 
 & \times 
 \biggl \{
 \frac{1}{\sinh{A}}
 \ln
 \biggl [ 
 \frac{\cosh{(A +B)} +\cosh{C}}{\cosh{(A -B)} +\cosh{C} }
 \biggr ] 
 \biggr \}^2 . 
 \label{eq61}
\end{align}
 
In the high temperature regime, the Fermi distribution function reduces to the Maxwell-Boltzmann distribution function. 
In this case, the function $F$ reads 
\begin{align}
 F \simeq &
 \tilde{z}_{\uparrow}\tilde{z}_{\downarrow} \exp \left [ -\tilde{\beta} \left ( \frac{\boldsymbol{P}^2}{2M} +\frac{\boldsymbol{p}^2}{2 m_{\rm re} } \right ) \right ] . 
\end{align} 
 As a result, we have the following form for the relaxation time:
 \begin{align}
 \frac{N_{\text{tot}}}{V}
 \frac{1}{\tau }
 \simeq &
 \frac{g^2  m_{\rm re}^3
 M^{3/2}}{2^{3/2} \pi^{9/2} \hbar^{10}}
 \tilde{z}_{\uparrow}\tilde{z}_{\downarrow}
 (k_B \tilde{T})^{7/2}. 
\label{eq63}
\end{align}
If the form (\ref{eq63}) does not reproduce well the temperature dependence of (\ref{eq61}) in numerical calculations, one needs the higher order fugacity expansion.

\section{conservation laws in moment equation}\label{AppendixB}

In this appendix, we summarize the way to determine the coefficients $\tilde a_{\sigma}$, $\tilde b_{\sigma} \equiv | \tilde {\bf b}_{\sigma}|$, and $\tilde c_{\sigma}$ in Eq. (\ref{eq:20}). 
These coefficients are determined by the conservation law of number of particles, momentum, as well as energy, the condition of which are respectively given by 
    \begin{align}
 	\langle \nu_{\sigma,0} \rangle
 	-a_{\sigma}W_{\sigma,0} -c_{\sigma}W_{\sigma,2}=0, 
	\\
 	\sum_{\sigma}
 	\bigl(
 	\langle p \nu_{\sigma,1} \rangle -b_{\sigma} W_{\sigma,2}
 	\bigr)=0, 
	\\
 	\sum_{\sigma} \frac{1}{m_{\sigma}}\bigl(
 	\langle p^2 \nu_{\sigma,0} \rangle
 	-a_{\sigma} W_{\sigma,2} -c_{\sigma} W_{\sigma,4}
 	\bigr)=0. 
    \end{align}
Because of local equilibrium conditions $\delta \tilde \theta_{\sigma} = \delta \tilde \theta$, $\delta \tilde {\bf v}_{\sigma} =\delta \tilde {\bf v}$ as well as a static equilibrium condition $\beta_{\sigma}^0 = \beta^0$, we have also employed conditions $\tilde b_{\uparrow} = \tilde b_{\downarrow} \equiv \tilde b$ as well as $m_{\uparrow} \tilde c_{\uparrow} = m_{\downarrow} \tilde c_{\downarrow}$.  
Using these conditions, we have determined the coefficients 
    \begin{align}
 	\tilde a_{\sigma}
 	= & \frac{1}{W_{\sigma,0}} \langle \nu_{\sigma,0} \rangle
	 	\label{eq:coefficient-A}
	\\ & 
 	-\frac{1}{m_{\sigma} \Phi}\frac{W_{\sigma,2}}{W_{\sigma,0}}
 	\sum_{\sigma'} \frac{1}{ m_{\sigma'}}
	\left ( 
	\langle p^2 \nu_{\sigma',0} \rangle
 	-  
 	\frac{W_{\sigma',2}}{W_{\sigma',0}}
 	\langle \nu_{\sigma',0} \rangle
	\right ) , 
\nonumber 
 	\\
 	\tilde b
 	= & \frac{
 	 \sum_{\sigma} \langle p \nu_{\sigma,1} \rangle 
 	}{
 	\sum_{\sigma} W_{\sigma,2} 
 	},   
 	\label{eq:coefficient-B}
 	\\
 	\tilde c_{\sigma}
 	= & \frac{1}{m_{\sigma} \Phi}
 	\sum_{\sigma'} \frac{1}{ m_{\sigma'}}
	\left ( 
	\langle p^2 \nu_{\sigma',0} \rangle
 	-  
 	\frac{W_{\sigma',2}}{W_{\sigma',0}}
 	\langle \nu_{\sigma',0} \rangle
	\right )
	. 
 	\label{eq:coefficient-C}  
    \end{align} 
As a result, we obtain the moment equation (\ref{eq:moment-eq}). 

\section{first sound}\label{AppendixC}

In this Appendix, we discuss collective modes in the hydrodynamic regime, assuming $U_{{\rm ext},\sigma}=0$.
Using the expression (\ref{eq:dtf}), we find that the density fluctuation is given in the form 
 \begin{align}
  \delta n_{\sigma}({\bf q},\omega)
  =& \tilde a_{\sigma} W_{\sigma , 0} + \tilde c_{\sigma} W_{\sigma , 2}
  \nonumber \\ 
 &   +\int \frac{d{\bf p}}{(2 \pi \hbar)^3} \frac{\partial f^0_{\sigma}}{\partial \epsilon^0_{\sigma}}\delta \nu_{\sigma}({\bf p}, {\bf q},\omega), 
  \label{eq:17}
 \end{align}
where $\delta \nu ({\bf p}, {\bf q},\omega) \equiv \nu_{\sigma}({\bf p}, {\bf q},\omega) - \tilde \nu_{\sigma}({\bf p}, {\bf q},\omega)$ represents 
the deviation from the local equilibrium. 
We here consider the approximation form of $\delta \nu_{\sigma}$ in the hydrodynamic regime by using the facts that the deviation from the local equilibrium is very small, and the relaxation time is short in that regime. 
By expanding $\delta \nu_{\sigma}$ with the parameter $\tau$  small in the hydrodynamic regime, and applying the expanded form of $\delta \nu_{\sigma}$ to (\ref{eq:20}), 
we find the lowest order solution of $\delta \nu_{\sigma}$ with respect to $\tau$ given in the form 
\begin{align}
 \delta \nu_{\sigma}({\bf p}, {\bf q},\omega)
 = & i \tau 
\biggl [
 \left (\omega-\frac{{\bf p} \cdot {\bf q}}{m_{\sigma}} \right )
( \tilde a_{\sigma}  +  \tilde{\bf b}  \cdot {\bf p} +  \tilde c_{\sigma} p^2  )
\nonumber
 \\
 & 
 +
 \frac{{\bf p} \cdot {\bf q}}{m_{\sigma}}
 g 
 \left (
 \tilde{a}_{-\sigma} W_{-\sigma , 0}
 +\tilde{c}_{-\sigma} W_{-\sigma , 2}
 \right )
\biggr ], 
 \label{eq:b1}
\end{align}
where we have also used the local equilibrium condition $\tilde {\bf b}_{\uparrow} = \tilde {\bf b}_{\downarrow}$. 
We eliminate $\omega$-dependence from (\ref{eq:b1}) by using the lowest-order hydrodynamic equations with respect to $\tau$. 
As a result, we have a form of the fluctuation $  \delta \nu_{\sigma}$, given by  
\begin{align}
  \delta \nu_{\sigma}({\bf p}, {\bf q},\omega)
 =&i \tau 
 \biggl [ 
 \frac{{\bf p} \cdot {\bf q}}{2m_{\sigma}^2}\beta^0
 ( 
 p^2 - \mathcal{J}_{\sigma}
 )
 \delta \tilde{\theta} 
 \nonumber 
 \\ 
 & 
 +\biggl (
 -\frac{\delta \tilde{{\bf v}} 
 \cdot {\bf q}}{3 m_{\sigma}} p^2
 +\frac{{\bf p} \cdot {\bf q}}{m_{\sigma}}
  \delta \tilde{{\bf v}}  \cdot {\bf p}
 \biggr ) 
 \biggr ], 
 \label{eq:b7}
\end{align}
where 
\begin{align}
  \mathcal{J}_{\sigma}
  \equiv& \frac{1}{\sum_{\sigma'} W_{\sigma',2}}
  \biggl [ 
m_{\sigma}^2
  \sum_{\sigma'} 
  \frac{W_{\sigma',4} }{m_{\sigma'}^2} 
 \nonumber\\
  &
  -  \frac{g m_{\sigma}}{m_{-\sigma}}
  W_{-\sigma,2} ( W_{\sigma , 2} - W_{-\sigma , 2} ) 
  \biggr ] , 
\end{align} 
and we have used the relation $m_{\uparrow} \tilde c_{\uparrow} = m_{\downarrow} \tilde c_{\downarrow}$. 
The form (\ref{eq:b7}) satisfies the conservation laws 
 \begin{align}
  \int \frac{d{\bf p}}{(2 \pi \hbar)^3}
  \frac{\partial f^0_{\sigma}}{\partial \epsilon^0_{\sigma}}
  \delta \nu_{\sigma}({\bf p}, {\bf q},\omega)
  = 0, 
  \\
  \sum_{\sigma}
  \int \frac{d{\bf p}}{(2 \pi \hbar)^3}
  \frac{\partial f^0_{\sigma}}{\partial \epsilon^0_{\sigma}}
  {\bf p}
  \delta \nu_{\sigma}({\bf p}, {\bf q},\omega)
  = 0,  
  \\
  \sum_{\sigma}
  \int \frac{d{\bf p}}{(2 \pi \hbar)^3}
  \frac{\partial f^0_{\sigma}}{\partial \epsilon^0_{\sigma}}
  \frac{p^2}{m_{\sigma}} 
  \delta \nu_{\sigma}({\bf p}, {\bf q},\omega)
  = 0. 
 \end{align}

We then derive the hydrodynamic equation by using the form $\nu_{\sigma} ({\bf p}) = \tilde a_{\sigma}+ \tilde {\bf b}_{ } \cdot {\bf p} + \tilde c_{\sigma}p^2 +   \delta \nu_{\sigma}({\bf p})$, which of density, momentum, as well as energy are respectively given by  
\begin{widetext}
\begin{align}
 \omega \tilde a_{\sigma} W_{\sigma,0}
  +\omega \tilde c_{\sigma} W_{\sigma,2}
  -\frac{\tilde{\bf b} \cdot {\bf q}}{3m_{\sigma}}  W_{\sigma,2}
  +
  i \tau \tilde  c_{\sigma} \frac{q^2}{3}
  \mathcal{G}_{\sigma}
 & = 0, 
 \label{eq:fs-1}
 \\ 
 \frac{1}{3}
 \sum_{\sigma}
  \left [ 
  \omega W_{\sigma,2} \tilde{\bf b}
  +\frac{{\bf q}}{m_{\sigma}} 
  (
  -\tilde a_{\sigma}  W_{\sigma,2}
  -\tilde c_{\sigma}  W_{\sigma,4}
 +g
 \tilde a_{-\sigma} W_{-\sigma,0} W_{\sigma , 2}
 +g
 \tilde c_{-\sigma} W_{\sigma,2} W_{-\sigma , 2}
 ) 
 \right ] & 
 \nonumber
 \\ 
	- \frac{1}{3} i \omega
	{\bf q}
		\tau 
	\sum_{\sigma}
	m_{\sigma} \tilde c_{\sigma}
	\mathcal{G}_{\sigma} 
	+
	i
		\tau  
	\sum_{\sigma}
	\frac{W_{\sigma,4}}{15 m_{\sigma}^2} 
	\Bigl[
	q^2
	 \tilde{\bf b}
	+
	\frac{1}{3}
	{\bf q}
	(  \tilde{\bf b} \cdot {\bf q})
	\Bigr] 
 & = 0, 
 \label{eq:fs-2}
 \\ 
 \sum_{\sigma}
 \frac{1}{m_{\sigma}}
\left [ 
 \omega \tilde a_{\sigma} W_{\sigma,2}
 +\omega \tilde c_{\sigma} W_{\sigma,4}
 -\frac{\tilde{\bf b} \cdot {\bf q}}{3m_{\sigma}}  W_{\sigma,4}
 +i \tau 
 \frac{q^2}{6m_{\sigma}^2}
 ( 
 W_{\sigma,6} -  W_{\sigma,4}\mathcal{J}_{\sigma}
) 
 \tilde c_{\sigma}
\right ] 
 & = 0, 
 \label{eq:fs-3}
\end{align}
\end{widetext}
where $ \mathcal{G}_{\sigma} \equiv (W_{\sigma,4} - W_{\sigma,2} \mathcal{J}_{\sigma}) / m_{\sigma}^2$. 
The coefficients $\tilde a_{\sigma}$, $\tilde {\bf b}$, and $\tilde c_{\sigma}$ are related to the hydrodynamic variables: 
the density fluctuation $\delta \tilde n_{\sigma}  = \tilde a_{\sigma} W_{\sigma , 0} + \tilde c_{\sigma} W_{\sigma , 2}$, 
the velocity fluctuation $\tilde{{\bf v}} = -\tilde {\bf b}$, as well as the energy fluctuation $  \delta \tilde E = \sum_{\sigma} [ \tilde a_{\sigma} W_{\sigma,2} + \tilde c_{\sigma} W_{\sigma,4}]/(2m_{\sigma})$. 
The equations (\ref{eq:fs-1}), (\ref{eq:fs-2}), and (\ref{eq:fs-3}) are eigenvalue equations in the hydrodynamic regime,
and an eigenvalue $\omega \equiv \Omega - i \Gamma$ provides the frequency of the hydrodynamic mode $\Omega$ as well as its damping rate $\Gamma$, where $\Omega$ ($\Gamma$) is a real (imaginary) part of $\omega$. 
In the hydrodynamic regime where $ | \omega |  \tau \ll 1$ holds, 
the eigenvalue equations (\ref{eq:fs-1})-(\ref{eq:fs-3}) are approximately reduced into the form 
 \begin{align}
 0 = & F_1(\omega) +i \tau F_2(\omega),
 \label{eq:F1F2} 
 \end{align}
where we have omitted the terms of second and higher order in $\tau$, and $F_1$ and $F_2$ are real functions that take the form
\begin{align}
F_1(\omega)=\omega^2 ( \omega ^2 -\Omega_{\rm HD}^{2}),~~
F_2(\omega)=\omega(B \omega^2 -C) . 
\label{eq:F1}
\end{align} 
In Eq.~(\ref{eq:F1}), $\Omega_{\rm HD}$ is the non-zero solution of $F_1(\omega)=0$, which is given by
 \begin{align} 
\Omega_{\rm HD}^2=
\frac{1}{3\sum_{\sigma} W_{\sigma,2}}
 \biggl(
 \Bigl[
 \sum_{\sigma} \frac{W_{\sigma,4}}{m_{\sigma}^2}
 \Bigr]
 -
2
 g 
 \frac{W_{\uparrow,2} W_{\downarrow,2}}{m_{\uparrow} m_{\downarrow}}
 \biggr)q^2 . 
 \label{fsomega2}
 \end{align}
Although we do not give explicit expressions for coefficients $B$, and $C$, those are given in terms of $W_{\sigma, n}$, $m_{\sigma}$, and so on.  
 In the hydrodynamic regime, the frequency $\Omega$ is given by $\Omega\approx \Omega_{\rm HD}$.
 The damping rate is obtained by substituting $\omega=\Omega_{\rm HD}-i\Gamma$ into (\ref{eq:F1F2}) and
 using the condition $\Omega_{\rm HD} \gg \Gamma$. Expanding the solution for $\Gamma$ to first order in $\tau$,  we obtain
 \begin{align}
 \Gamma
  =&
  \frac{B \Omega_{\rm HD}^2  -C}{2 \Omega_{\rm HD}^2} \tau , 
 \end{align}
 which is proportional to the relaxation time $\tau$.

\section{zero sound}\label{AppendixD}
In this Appendix, we consider the collective mode in the collisionless regime at an extremely low temperature $T \ll {\rm min} (T_{{\rm F}, \uparrow}, T_{{\rm F}, \downarrow} )$, by extending the study for the zero sound in the mass and population balanced gas~\cite{Khalatnikov1958,Larionov1999,Watabe2009}. 
In this case, from (\ref{eq:coefficient-A}), (\ref{eq:coefficient-B}), and (\ref{eq:coefficient-C}), one can take $\tilde a_{\sigma} = \langle \nu_{\sigma,0} \rangle / W_{\sigma,0}$, $| \tilde {\bf b}| = \sum_{\sigma} \langle p \nu_{\sigma,1} \rangle / \sum_{\sigma} W_{\sigma,2}$ as well as $\tilde c_{\sigma} = 0$, because of relations at $T = 0$, given by 
 \begin{align}
  W_{\sigma,n}  
  &=
  -
  \frac{4 \pi}{(2\pi \hbar)^3} \frac{1}{v_{F,\sigma}}
  p_{F,\sigma}^{n+2}, 
  \\
  \langle p^n\nu_{\sigma,l} \rangle
  =&
  \frac{4 \pi}{(2\pi \hbar)^3} \frac{1}{v_{F,\sigma}}
  p_{F,\sigma}^{n+2} \nu_{\sigma,l}(p_{F,\sigma}). 
 \end{align} 
Applying these results to (\ref{eq:20}), one can rewrite the linearized Boltzmann equation (with $U_{{\rm ext},\sigma}=0$) as 
\begin{align}
 & \frac{\partial f^0_{\sigma}}{\partial \epsilon^0_{\sigma}}
 \Biggl[
 \sum_{l=0}^{\infty}  \nu_{\sigma,l}(p) P_l(\cos \theta)
 -
 \cfrac{\cos \theta}{\cos \theta - \cfrac{i \tau  \omega -1}{i \tau p q /m_{\sigma}}}
 +g  \langle \nu_{-\sigma,0} \rangle 
 \Biggr]
 \nonumber
 \\
 = & 
 \frac{1}{i \tau}
 \frac{\partial f^0_{\sigma}}{\partial \epsilon^0_{\sigma}}
 \frac{1}{pq /m_{\sigma}}
 \cfrac{1}{\cos \theta - \cfrac{i \tau  \omega -1}{i \tau p q /m_{\sigma}}}
 \biggl[
 \frac{\langle \nu_{\sigma,0} \rangle}{W_{\sigma,0}}
 +
 \frac{ \sum_{\sigma} \langle p \nu_{\sigma,1} \rangle }
 { \sum_{\sigma} W_{\sigma,2} }
 p \cos \theta
 \biggr]. 
 \label{eq:zero50}
\end{align}
Multiplying $1$ and $p \cos \theta$ to (\ref{eq:zero50}) and integrating over ${\bf p}$, we end up with 
\begin{widetext}
\begin{align}
 \frac{1}{A^{\sigma}_{0,1}}
 \biggl(
 1 
 -\frac{1}{i \tau  q /m_{\sigma}} \frac{A_{-1,0}^{\sigma}}{W_{\sigma,0}} 
 \biggr)
 \langle \nu_{\sigma,0} \rangle
 -
 g 
 \langle \nu_{-\sigma,0} \rangle
 -
 \frac{1}{i \tau q /m_{\sigma}}
 \frac{1}{\sum_{\sigma} W_{\sigma,2} }
 \sum_{\sigma} \langle p \nu_{\sigma,1} \rangle 
 & =  
0, 
\label{eq:z1}
 \\ 
 \frac{ 1}{A^{\sigma}_{1,2}}
 \biggl ( 
 \frac{1}{i \tau  q /m_{\sigma}} \frac{A_{0,1}^{\sigma}}{W_{\sigma,0}} 
 \biggr)
 \langle \nu_{\sigma,0} \rangle
+ 
 g 
 \langle \nu_{-\sigma,0} \rangle 
 - 
 \frac{1}{3 A^{\sigma}_{1,2}}
 \langle p \nu_{\sigma,1} \rangle
 +
 \frac{1}{i \tau q /m_{\sigma}}
 \frac{1}{\sum_{\sigma} W_{\sigma,2}  }
 \sum_{\sigma} \langle p \nu_{\sigma,1} \rangle 
 & = 
0, 
\label{eq:z2}
\end{align}
\end{widetext}
where 
\begin{align}
 A^{\sigma}_{n,l}
 \equiv&
 \int
 \frac{d {\bf p}}{(2\pi \hbar)^3}
 \frac{\partial f^0_{\sigma}}{\partial \epsilon ^0_{\sigma}}
 p^n
 \cfrac{\cos^l \theta}{\cos \theta - \cfrac{i \tau \omega -1}{i \tau p q /m_{\sigma}}} . 
\end{align}
Taking the zero temperature limit for $f_{\sigma}^{0}$, one finds
\begin{align}
 A^{\sigma}_{n,0}
 = &
C_{n}^{\sigma}
 \ln \left | \frac{s_{\sigma}+1}{s_{\sigma} -1} \right |, 
 \\
 A^{\sigma}_{n,1}
 =&
 -
C_{n}^{\sigma}
 \biggl ( 
 2 - s_{\sigma}\ln \left | \frac{s_{\sigma}+1}{s_{\sigma} -1} \right |
 \biggr ), 
  \\
 A^{\sigma}_{n,2}
 =&
 -
C_{n}^{\sigma}
s_{\sigma}
 \biggl ( 
 2 - s_{\sigma}  \ln \left | \frac{s_{\sigma}+1}{s_{\sigma} -1} \right |
 \biggr ), 
\end{align}
with $  s_{\sigma} \equiv (i \tau  \omega -1 )/ (i \tau  v_{F,\sigma} q)$ and $C_{n}^{\sigma} \equiv (mv_{F,\sigma})^{n+2} / ( 4 \pi^{2} \hbar^{3} v_{F,\sigma})$. 
The zero sound velocity is obtained by solving (\ref{eq:z1}) and (\ref{eq:z2}) to find the non-trivial solution of $\langle \nu_{\sigma,0} \rangle$ as well as $\langle p \nu_{\sigma,1} \rangle$. 
In particular, close to the zero temperature, the mean-collision time drastically increases because of the Pauli blocking, and we may take $\tau \rightarrow \infty$, which leads to $s_{\sigma} \rightarrow \Omega /(v_{f,\sigma} q)$ as well as $1/ (i \tau q / m_{\sigma} )  \rightarrow 0$. 
As a result, we have the equation to determine the zero sound mode frequency $\Omega$, given by 
\begin{align}
 1 - g^{2} A_{0,1}^{\uparrow} A_{0,1}^{\downarrow} 
= 0. 
\end{align}

\bibliographystyle{apsrev4-1}
\bibliography{library.bib}

\begin{thebibliography}{39}%
\makeatletter
\providecommand \@ifxundefined [1]{%
 \@ifx{#1\undefined}
}%
\providecommand \@ifnum [1]{%
 \ifnum #1\expandafter \@firstoftwo
 \else \expandafter \@secondoftwo
 \fi
}%
\providecommand \@ifx [1]{%
 \ifx #1\expandafter \@firstoftwo
 \else \expandafter \@secondoftwo
 \fi
}%
\providecommand \natexlab [1]{#1}%
\providecommand \enquote  [1]{``#1''}%
\providecommand \bibnamefont  [1]{#1}%
\providecommand \bibfnamefont [1]{#1}%
\providecommand \citenamefont [1]{#1}%
\providecommand \href@noop [0]{\@secondoftwo}%
\providecommand \href [0]{\begingroup \@sanitize@url \@href}%
\providecommand \@href[1]{\@@startlink{#1}\@@href}%
\providecommand \@@href[1]{\endgroup#1\@@endlink}%
\providecommand \@sanitize@url [0]{\catcode `\\12\catcode `\$12\catcode
  `\&12\catcode `\#12\catcode `\^12\catcode `\_12\catcode `\%12\relax}%
\providecommand \@@startlink[1]{}%
\providecommand \@@endlink[0]{}%
\providecommand \url  [0]{\begingroup\@sanitize@url \@url }%
\providecommand \@url [1]{\endgroup\@href {#1}{\urlprefix }}%
\providecommand \urlprefix  [0]{URL }%
\providecommand \Eprint [0]{\href }%
\providecommand \doibase [0]{http://dx.doi.org/}%
\providecommand \selectlanguage [0]{\@gobble}%
\providecommand \bibinfo  [0]{\@secondoftwo}%
\providecommand \bibfield  [0]{\@secondoftwo}%
\providecommand \translation [1]{[#1]}%
\providecommand \BibitemOpen [0]{}%
\providecommand \bibitemStop [0]{}%
\providecommand \bibitemNoStop [0]{.\EOS\space}%
\providecommand \EOS [0]{\spacefactor3000\relax}%
\providecommand \BibitemShut  [1]{\csname bibitem#1\endcsname}%
\let\auto@bib@innerbib\@empty
\bibitem [{\citenamefont {Kinast}\ \emph {et~al.}(2004)\citenamefont {Kinast},
  \citenamefont {Hemmer}, \citenamefont {Gehm}, \citenamefont {Turlapov},\ and\
  \citenamefont {Thomas}}]{Kinast2004}%
  \BibitemOpen
  \bibfield  {author} {\bibinfo {author} {\bibfnamefont {J.}~\bibnamefont
  {Kinast}}, \bibinfo {author} {\bibfnamefont {S.~L.}\ \bibnamefont {Hemmer}},
  \bibinfo {author} {\bibfnamefont {M.~E.}\ \bibnamefont {Gehm}}, \bibinfo
  {author} {\bibfnamefont {A.}~\bibnamefont {Turlapov}}, \ and\ \bibinfo
  {author} {\bibfnamefont {J.~E.}\ \bibnamefont {Thomas}},\ }\href
  {https://link.aps.org/doi/10.1103/PhysRevLett.92.150402} {\bibfield
  {journal} {\bibinfo  {journal} {Physical Review Letters}\ }\textbf {\bibinfo
  {volume} {92}},\ \bibinfo {pages} {150402} (\bibinfo {year}
  {2004})}\BibitemShut {NoStop}%
\bibitem [{\citenamefont {Bartenstein}\ \emph {et~al.}(2004)\citenamefont
  {Bartenstein}, \citenamefont {Altmeyer}, \citenamefont {Riedl}, \citenamefont
  {Jochim}, \citenamefont {Chin}, \citenamefont {Denschlag},\ and\
  \citenamefont {Grimm}}]{Bartenstein2004}%
  \BibitemOpen
  \bibfield  {author} {\bibinfo {author} {\bibfnamefont {M.}~\bibnamefont
  {Bartenstein}}, \bibinfo {author} {\bibfnamefont {A.}~\bibnamefont
  {Altmeyer}}, \bibinfo {author} {\bibfnamefont {S.}~\bibnamefont {Riedl}},
  \bibinfo {author} {\bibfnamefont {S.}~\bibnamefont {Jochim}}, \bibinfo
  {author} {\bibfnamefont {C.}~\bibnamefont {Chin}}, \bibinfo {author}
  {\bibfnamefont {J.~H.}\ \bibnamefont {Denschlag}}, \ and\ \bibinfo {author}
  {\bibfnamefont {R.}~\bibnamefont {Grimm}},\ }\href
  {https://link.aps.org/doi/10.1103/PhysRevLett.92.203201} {\bibfield
  {journal} {\bibinfo  {journal} {Physical Review Letters}\ }\textbf {\bibinfo
  {volume} {92}},\ \bibinfo {pages} {203201} (\bibinfo {year}
  {2004})}\BibitemShut {NoStop}%
\bibitem [{\citenamefont {Altmeyer}\ \emph {et~al.}(2007)\citenamefont
  {Altmeyer}, \citenamefont {Riedl}, \citenamefont {Wright}, \citenamefont
  {Kohstall}, \citenamefont {Denschlag},\ and\ \citenamefont
  {Grimm}}]{Altmeyer2007}%
  \BibitemOpen
  \bibfield  {author} {\bibinfo {author} {\bibfnamefont {A.}~\bibnamefont
  {Altmeyer}}, \bibinfo {author} {\bibfnamefont {S.}~\bibnamefont {Riedl}},
  \bibinfo {author} {\bibfnamefont {M.~J.}\ \bibnamefont {Wright}}, \bibinfo
  {author} {\bibfnamefont {C.}~\bibnamefont {Kohstall}}, \bibinfo {author}
  {\bibfnamefont {J.~H.}\ \bibnamefont {Denschlag}}, \ and\ \bibinfo {author}
  {\bibfnamefont {R.}~\bibnamefont {Grimm}},\ }\href@noop {} {\bibfield
  {journal} {\bibinfo  {journal} {Physical Review A}\ }\textbf {\bibinfo
  {volume} {76}},\ \bibinfo {pages} {033610} (\bibinfo {year}
  {2007})}\BibitemShut {NoStop}%
\bibitem [{\citenamefont {Joseph}\ \emph {et~al.}(2007)\citenamefont {Joseph},
  \citenamefont {Clancy}, \citenamefont {Luo}, \citenamefont {Kinast},
  \citenamefont {Turlapov},\ and\ \citenamefont {Thomas}}]{Joseph2007}%
  \BibitemOpen
  \bibfield  {author} {\bibinfo {author} {\bibfnamefont {J.}~\bibnamefont
  {Joseph}}, \bibinfo {author} {\bibfnamefont {B.}~\bibnamefont {Clancy}},
  \bibinfo {author} {\bibfnamefont {L.}~\bibnamefont {Luo}}, \bibinfo {author}
  {\bibfnamefont {J.}~\bibnamefont {Kinast}}, \bibinfo {author} {\bibfnamefont
  {A.}~\bibnamefont {Turlapov}}, \ and\ \bibinfo {author} {\bibfnamefont
  {J.~E.}\ \bibnamefont {Thomas}},\ }\href
  {https://link.aps.org/doi/10.1103/PhysRevLett.98.170401} {\bibfield
  {journal} {\bibinfo  {journal} {Physical Review Letters}\ }\textbf {\bibinfo
  {volume} {98}},\ \bibinfo {pages} {170401} (\bibinfo {year}
  {2007})}\BibitemShut {NoStop}%
\bibitem [{\citenamefont {Wright}\ \emph {et~al.}(2007)\citenamefont {Wright},
  \citenamefont {Riedl}, \citenamefont {Altmeyer}, \citenamefont {Kohstall},
  \citenamefont {Guajardo}, \citenamefont {Denschlag},\ and\ \citenamefont
  {Grimm}}]{Wright2007}%
  \BibitemOpen
  \bibfield  {author} {\bibinfo {author} {\bibfnamefont {M.~J.}\ \bibnamefont
  {Wright}}, \bibinfo {author} {\bibfnamefont {S.}~\bibnamefont {Riedl}},
  \bibinfo {author} {\bibfnamefont {A.}~\bibnamefont {Altmeyer}}, \bibinfo
  {author} {\bibfnamefont {C.}~\bibnamefont {Kohstall}}, \bibinfo {author}
  {\bibfnamefont {E.~R.~S.}\ \bibnamefont {Guajardo}}, \bibinfo {author}
  {\bibfnamefont {J.~H.}\ \bibnamefont {Denschlag}}, \ and\ \bibinfo {author}
  {\bibfnamefont {R.}~\bibnamefont {Grimm}},\ }\href@noop {} {\bibfield
  {journal} {\bibinfo  {journal} {Physical Review Letters}\ }\textbf {\bibinfo
  {volume} {99}},\ \bibinfo {pages} {150403} (\bibinfo {year}
  {2007})}\BibitemShut {NoStop}%
\bibitem [{\citenamefont {Tey}\ \emph {et~al.}(2013)\citenamefont {Tey},
  \citenamefont {Sidorenkov}, \citenamefont {Guajardo}, \citenamefont {Grimm},
  \citenamefont {Ku}, \citenamefont {Zwierlein}, \citenamefont {Hou},
  \citenamefont {Pitaevskii},\ and\ \citenamefont {Stringari}}]{Tey2013}%
  \BibitemOpen
  \bibfield  {author} {\bibinfo {author} {\bibfnamefont {M.~K.}\ \bibnamefont
  {Tey}}, \bibinfo {author} {\bibfnamefont {L.~A.}\ \bibnamefont {Sidorenkov}},
  \bibinfo {author} {\bibfnamefont {E.~R.~S.}\ \bibnamefont {Guajardo}},
  \bibinfo {author} {\bibfnamefont {R.}~\bibnamefont {Grimm}}, \bibinfo
  {author} {\bibfnamefont {M.~J.~H.}\ \bibnamefont {Ku}}, \bibinfo {author}
  {\bibfnamefont {M.~W.}\ \bibnamefont {Zwierlein}}, \bibinfo {author}
  {\bibfnamefont {Y.-H.}\ \bibnamefont {Hou}}, \bibinfo {author} {\bibfnamefont
  {L.}~\bibnamefont {Pitaevskii}}, \ and\ \bibinfo {author} {\bibfnamefont
  {S.}~\bibnamefont {Stringari}},\ }\href
  {http://link.aps.org/doi/10.1103/PhysRevLett.110.055303} {\bibfield
  {journal} {\bibinfo  {journal} {Physical Review Letters}\ }\textbf {\bibinfo
  {volume} {110}},\ \bibinfo {pages} {55303} (\bibinfo {year}
  {2013})}\BibitemShut {NoStop}%
\bibitem [{\citenamefont {Sidorenkov}\ \emph {et~al.}(2013)\citenamefont
  {Sidorenkov}, \citenamefont {Tey}, \citenamefont {Grimm}, \citenamefont
  {Hou}, \citenamefont {Pitaevskii},\ and\ \citenamefont
  {Stringari}}]{Sidorenkov2013}%
  \BibitemOpen
  \bibfield  {author} {\bibinfo {author} {\bibfnamefont {L.~A.}\ \bibnamefont
  {Sidorenkov}}, \bibinfo {author} {\bibfnamefont {M.~K.}\ \bibnamefont {Tey}},
  \bibinfo {author} {\bibfnamefont {R.}~\bibnamefont {Grimm}}, \bibinfo
  {author} {\bibfnamefont {Y.-H.}\ \bibnamefont {Hou}}, \bibinfo {author}
  {\bibfnamefont {L.}~\bibnamefont {Pitaevskii}}, \ and\ \bibinfo {author}
  {\bibfnamefont {S.}~\bibnamefont {Stringari}},\ }\href {\doibase
  10.1038/nature12136
  http://www.nature.com/nature/journal/v498/n7452/abs/nature12136.html#supplementary-information}
  {\bibfield  {journal} {\bibinfo  {journal} {Nature}\ }\textbf {\bibinfo
  {volume} {498}},\ \bibinfo {pages} {78} (\bibinfo {year} {2013})}\BibitemShut
  {NoStop}%
\bibitem [{\citenamefont {Gensemer}\ and\ \citenamefont
  {Jin}(2001)}]{Gensemer2001}%
  \BibitemOpen
  \bibfield  {author} {\bibinfo {author} {\bibfnamefont {S.~D.}\ \bibnamefont
  {Gensemer}}\ and\ \bibinfo {author} {\bibfnamefont {D.~S.}\ \bibnamefont
  {Jin}},\ }\href {https://link.aps.org/doi/10.1103/PhysRevLett.87.173201}
  {\bibfield  {journal} {\bibinfo  {journal} {Physical Review Letters}\
  }\textbf {\bibinfo {volume} {87}},\ \bibinfo {pages} {173201} (\bibinfo
  {year} {2001})}\BibitemShut {NoStop}%
\bibitem [{\citenamefont {DeMarco}\ and\ \citenamefont
  {Jin}(2002)}]{DeMarco2002}%
  \BibitemOpen
  \bibfield  {author} {\bibinfo {author} {\bibfnamefont {B.}~\bibnamefont
  {DeMarco}}\ and\ \bibinfo {author} {\bibfnamefont {D.~S.}\ \bibnamefont
  {Jin}},\ }\href {https://link.aps.org/doi/10.1103/PhysRevLett.88.040405}
  {\bibfield  {journal} {\bibinfo  {journal} {Physical Review Letters}\
  }\textbf {\bibinfo {volume} {88}},\ \bibinfo {pages} {40405} (\bibinfo {year}
  {2002})}\BibitemShut {NoStop}%
\bibitem [{\citenamefont {Landau}(1957)}]{Landau1957}%
  \BibitemOpen
  \bibfield  {author} {\bibinfo {author} {\bibfnamefont {L.~D.}\ \bibnamefont
  {Landau}},\ }\href@noop {} {\bibfield  {journal} {\bibinfo  {journal} {Sov.
  Phys. JETP}\ }\textbf {\bibinfo {volume} {5}},\ \bibinfo {pages} {101}
  (\bibinfo {year} {1957})}\BibitemShut {NoStop}%
\bibitem [{\citenamefont {Abel}\ \emph {et~al.}(1966)\citenamefont {Abel},
  \citenamefont {Anderson},\ and\ \citenamefont {Wheatley}}]{Abel1966}%
  \BibitemOpen
  \bibfield  {author} {\bibinfo {author} {\bibfnamefont {W.~R.}\ \bibnamefont
  {Abel}}, \bibinfo {author} {\bibfnamefont {A.~C.}\ \bibnamefont {Anderson}},
  \ and\ \bibinfo {author} {\bibfnamefont {J.~C.}\ \bibnamefont {Wheatley}},\
  }\href {http://link.aps.org/doi/10.1103/PhysRevLett.17.74} {\bibfield
  {journal} {\bibinfo  {journal} {Physical Review Letters}\ }\textbf {\bibinfo
  {volume} {17}},\ \bibinfo {pages} {74} (\bibinfo {year} {1966})}\BibitemShut
  {NoStop}%
\bibitem [{\citenamefont {Nitsch}(1972)}]{Nitsch1972}%
  \BibitemOpen
  \bibfield  {author} {\bibinfo {author} {\bibfnamefont {J.}~\bibnamefont
  {Nitsch}},\ }\href@noop {} {\bibfield  {journal} {\bibinfo  {journal}
  {Zeitschrift f\"ur Physik A Hadrons and nuclei}\ }\textbf {\bibinfo {volume}
  {251}},\ \bibinfo {pages} {141} (\bibinfo {year} {1972})}\BibitemShut
  {NoStop}%
\bibitem [{\citenamefont {Kouvaris}(2009)}]{Kouvaris2009}%
  \BibitemOpen
  \bibfield  {author} {\bibinfo {author} {\bibfnamefont {C.}~\bibnamefont
  {Kouvaris}},\ }\href@noop {} {\bibfield  {journal} {\bibinfo  {journal}
  {Physical Review D}\ }\textbf {\bibinfo {volume} {79}},\ \bibinfo {pages}
  {123008} (\bibinfo {year} {2009})}\BibitemShut {NoStop}%
\bibitem [{\citenamefont {Vichi}(2000)}]{Vichi2000}%
  \BibitemOpen
  \bibfield  {author} {\bibinfo {author} {\bibfnamefont {L.}~\bibnamefont
  {Vichi}},\ }\href {\doibase 10.1023/A:1004815907236} {\bibfield  {journal}
  {\bibinfo  {journal} {Journal of Low Temperature Physics}\ }\textbf {\bibinfo
  {volume} {121}},\ \bibinfo {pages} {177} (\bibinfo {year}
  {2000})}\BibitemShut {NoStop}%
\bibitem [{\citenamefont {Khalatnikov}\ and\ \citenamefont
  {Abrikosov}(1958)}]{Khalatnikov1958}%
  \BibitemOpen
  \bibfield  {author} {\bibinfo {author} {\bibfnamefont {I.~M.}\ \bibnamefont
  {Khalatnikov}}\ and\ \bibinfo {author} {\bibfnamefont {A.~A.}\ \bibnamefont
  {Abrikosov}},\ }\href@noop {} {\bibfield  {journal} {\bibinfo  {journal}
  {Sov. Phys. JETP}\ }\textbf {\bibinfo {volume} {6}},\ \bibinfo {pages} {84}
  (\bibinfo {year} {1958})}\BibitemShut {NoStop}%
\bibitem [{\citenamefont {Larionov}\ \emph {et~al.}(1999)\citenamefont
  {Larionov}, \citenamefont {Cabibbo}, \citenamefont {Baran},\ and\
  \citenamefont {{Di Toro}}}]{Larionov1999}%
  \BibitemOpen
  \bibfield  {author} {\bibinfo {author} {\bibfnamefont {A.~B.}\ \bibnamefont
  {Larionov}}, \bibinfo {author} {\bibfnamefont {M.}~\bibnamefont {Cabibbo}},
  \bibinfo {author} {\bibfnamefont {V.}~\bibnamefont {Baran}}, \ and\ \bibinfo
  {author} {\bibfnamefont {M.}~\bibnamefont {{Di Toro}}},\ }\href {\doibase
  http://dx.doi.org/10.1016/S0375-9474(99)00027-5} {\bibfield  {journal}
  {\bibinfo  {journal} {Nuclear Physics A}\ }\textbf {\bibinfo {volume}
  {648}},\ \bibinfo {pages} {157} (\bibinfo {year} {1999})}\BibitemShut
  {NoStop}%
\bibitem [{\citenamefont {Watabe}\ \emph {et~al.}(2009)\citenamefont {Watabe},
  \citenamefont {Osawa},\ and\ \citenamefont {Nikuni}}]{Watabe2009}%
  \BibitemOpen
  \bibfield  {author} {\bibinfo {author} {\bibfnamefont {S.}~\bibnamefont
  {Watabe}}, \bibinfo {author} {\bibfnamefont {A.}~\bibnamefont {Osawa}}, \
  and\ \bibinfo {author} {\bibfnamefont {T.}~\bibnamefont {Nikuni}},\ }\href
  {\doibase 10.1007/s10909-009-0043-4} {\bibfield  {journal} {\bibinfo
  {journal} {Journal of Low Temperature Physics}\ }\textbf {\bibinfo {volume}
  {158}},\ \bibinfo {pages} {773} (\bibinfo {year} {2009})}\BibitemShut
  {NoStop}%
\bibitem [{\citenamefont {Watabe}\ and\ \citenamefont
  {Nikuni}(2010)}]{Watabe2010}%
  \BibitemOpen
  \bibfield  {author} {\bibinfo {author} {\bibfnamefont {S.}~\bibnamefont
  {Watabe}}\ and\ \bibinfo {author} {\bibfnamefont {T.}~\bibnamefont
  {Nikuni}},\ }\href {\doibase 10.1103/PhysRevA.82.033622} {\bibfield
  {journal} {\bibinfo  {journal} {Physical Review A}\ }\textbf {\bibinfo
  {volume} {82}} (\bibinfo {year} {2010}),\
  10.1103/PhysRevA.82.033622}\BibitemShut {NoStop}%
\bibitem [{\citenamefont {Ronen}\ and\ \citenamefont {Bohn}(2010)}]{Ronen2010}%
  \BibitemOpen
  \bibfield  {author} {\bibinfo {author} {\bibfnamefont {S.}~\bibnamefont
  {Ronen}}\ and\ \bibinfo {author} {\bibfnamefont {J.~L.}\ \bibnamefont
  {Bohn}},\ }\href {http://link.aps.org/doi/10.1103/PhysRevA.81.033601}
  {\bibfield  {journal} {\bibinfo  {journal} {Physical Review A}\ }\textbf
  {\bibinfo {volume} {81}},\ \bibinfo {pages} {33601} (\bibinfo {year}
  {2010})}\BibitemShut {NoStop}%
\bibitem [{\citenamefont {Li}\ \emph {et~al.}(2013)\citenamefont {Li},
  \citenamefont {Anderson},\ and\ \citenamefont {Miller}}]{Li2013}%
  \BibitemOpen
  \bibfield  {author} {\bibinfo {author} {\bibfnamefont {D.~Z.}\ \bibnamefont
  {Li}}, \bibinfo {author} {\bibfnamefont {R.~H.}\ \bibnamefont {Anderson}}, \
  and\ \bibinfo {author} {\bibfnamefont {M.~D.}\ \bibnamefont {Miller}},\
  }\href {http://link.aps.org/doi/10.1103/PhysRevB.87.104519} {\bibfield
  {journal} {\bibinfo  {journal} {Physical Review B}\ }\textbf {\bibinfo
  {volume} {87}},\ \bibinfo {pages} {104519} (\bibinfo {year}
  {2013})}\BibitemShut {NoStop}%
\bibitem [{\citenamefont {Yip}\ and\ \citenamefont {Ho}(1999)}]{Yip1999}%
  \BibitemOpen
  \bibfield  {author} {\bibinfo {author} {\bibfnamefont {S.~K.}\ \bibnamefont
  {Yip}}\ and\ \bibinfo {author} {\bibfnamefont {T.-L.}\ \bibnamefont {Ho}},\
  }\href {https://link.aps.org/doi/10.1103/PhysRevA.59.4653} {\bibfield
  {journal} {\bibinfo  {journal} {Physical Review A}\ }\textbf {\bibinfo
  {volume} {59}},\ \bibinfo {pages} {4653} (\bibinfo {year}
  {1999})}\BibitemShut {NoStop}%
\bibitem [{\citenamefont {Trenkwalder}\ \emph {et~al.}(2011)\citenamefont
  {Trenkwalder}, \citenamefont {Kohstall}, \citenamefont {Zaccanti},
  \citenamefont {Naik}, \citenamefont {Sidorov}, \citenamefont {Schreck},\ and\
  \citenamefont {Grimm}}]{Trenkwalder2011}%
  \BibitemOpen
  \bibfield  {author} {\bibinfo {author} {\bibfnamefont {A.}~\bibnamefont
  {Trenkwalder}}, \bibinfo {author} {\bibfnamefont {C.}~\bibnamefont
  {Kohstall}}, \bibinfo {author} {\bibfnamefont {M.}~\bibnamefont {Zaccanti}},
  \bibinfo {author} {\bibfnamefont {D.}~\bibnamefont {Naik}}, \bibinfo {author}
  {\bibfnamefont {A.~I.}\ \bibnamefont {Sidorov}}, \bibinfo {author}
  {\bibfnamefont {F.}~\bibnamefont {Schreck}}, \ and\ \bibinfo {author}
  {\bibfnamefont {R.}~\bibnamefont {Grimm}},\ }\href
  {http://link.aps.org/doi/10.1103/PhysRevLett.106.115304} {\bibfield
  {journal} {\bibinfo  {journal} {Physical Review Letters}\ }\textbf {\bibinfo
  {volume} {106}},\ \bibinfo {pages} {115304} (\bibinfo {year}
  {2011})}\BibitemShut {NoStop}%
\bibitem [{\citenamefont {Spiegelhalder}\ \emph {et~al.}(2009)\citenamefont
  {Spiegelhalder}, \citenamefont {Trenkwalder}, \citenamefont {Naik},
  \citenamefont {Hendl}, \citenamefont {Schreck},\ and\ \citenamefont
  {Grimm}}]{Spiegelhalder2009}%
  \BibitemOpen
  \bibfield  {author} {\bibinfo {author} {\bibfnamefont {F.~M.}\ \bibnamefont
  {Spiegelhalder}}, \bibinfo {author} {\bibfnamefont {A.}~\bibnamefont
  {Trenkwalder}}, \bibinfo {author} {\bibfnamefont {D.}~\bibnamefont {Naik}},
  \bibinfo {author} {\bibfnamefont {G.}~\bibnamefont {Hendl}}, \bibinfo
  {author} {\bibfnamefont {F.}~\bibnamefont {Schreck}}, \ and\ \bibinfo
  {author} {\bibfnamefont {R.}~\bibnamefont {Grimm}},\ }\href
  {http://link.aps.org/doi/10.1103/PhysRevLett.103.223203} {\bibfield
  {journal} {\bibinfo  {journal} {Physical Review Letters}\ }\textbf {\bibinfo
  {volume} {103}},\ \bibinfo {pages} {223203} (\bibinfo {year}
  {2009})}\BibitemShut {NoStop}%
\bibitem [{\citenamefont {Tiecke}\ \emph {et~al.}(2010)\citenamefont {Tiecke},
  \citenamefont {Goosen}, \citenamefont {Ludewig}, \citenamefont {Gensemer},
  \citenamefont {Kraft}, \citenamefont {Kokkelmans},\ and\ \citenamefont
  {Walraven}}]{Tiecke2010}%
  \BibitemOpen
  \bibfield  {author} {\bibinfo {author} {\bibfnamefont {T.~G.}\ \bibnamefont
  {Tiecke}}, \bibinfo {author} {\bibfnamefont {M.~R.}\ \bibnamefont {Goosen}},
  \bibinfo {author} {\bibfnamefont {A.}~\bibnamefont {Ludewig}}, \bibinfo
  {author} {\bibfnamefont {S.~D.}\ \bibnamefont {Gensemer}}, \bibinfo {author}
  {\bibfnamefont {S.}~\bibnamefont {Kraft}}, \bibinfo {author} {\bibfnamefont
  {S.~J. J. M.~F.}\ \bibnamefont {Kokkelmans}}, \ and\ \bibinfo {author}
  {\bibfnamefont {J.~T.~M.}\ \bibnamefont {Walraven}},\ }\href
  {http://link.aps.org/doi/10.1103/PhysRevLett.104.053202} {\bibfield
  {journal} {\bibinfo  {journal} {Physical Review Letters}\ }\textbf {\bibinfo
  {volume} {104}},\ \bibinfo {pages} {53202} (\bibinfo {year}
  {2010})}\BibitemShut {NoStop}%
\bibitem [{\citenamefont {Stringari}(2009)}]{Stringari2009}%
  \BibitemOpen
  \bibfield  {author} {\bibinfo {author} {\bibfnamefont {S.}~\bibnamefont
  {Stringari}},\ }\href
  {http://link.aps.org/doi/10.1103/PhysRevLett.102.110406} {\bibfield
  {journal} {\bibinfo  {journal} {Physical Review Letters}\ }\textbf {\bibinfo
  {volume} {102}},\ \bibinfo {pages} {110406} (\bibinfo {year}
  {2009})}\BibitemShut {NoStop}%
\bibitem [{\citenamefont {Dahal}\ \emph {et~al.}(2008)\citenamefont {Dahal},
  \citenamefont {Gaudio}, \citenamefont {Feldmann},\ and\ \citenamefont
  {Bedell}}]{Dahal2008}%
  \BibitemOpen
  \bibfield  {author} {\bibinfo {author} {\bibfnamefont {H.~P.}\ \bibnamefont
  {Dahal}}, \bibinfo {author} {\bibfnamefont {S.}~\bibnamefont {Gaudio}},
  \bibinfo {author} {\bibfnamefont {J.~D.}\ \bibnamefont {Feldmann}}, \ and\
  \bibinfo {author} {\bibfnamefont {K.~S.}\ \bibnamefont {Bedell}},\ }\href
  {http://link.aps.org/doi/10.1103/PhysRevA.78.035601} {\bibfield  {journal}
  {\bibinfo  {journal} {Physical Review A}\ }\textbf {\bibinfo {volume} {78}},\
  \bibinfo {pages} {35601} (\bibinfo {year} {2008})}\BibitemShut {NoStop}%
\bibitem [{\citenamefont {Recati}\ and\ \citenamefont
  {Stringari}(2010)}]{Recati2010}%
  \BibitemOpen
  \bibfield  {author} {\bibinfo {author} {\bibfnamefont {A.}~\bibnamefont
  {Recati}}\ and\ \bibinfo {author} {\bibfnamefont {S.}~\bibnamefont
  {Stringari}},\ }\href {http://link.aps.org/doi/10.1103/PhysRevA.82.013635}
  {\bibfield  {journal} {\bibinfo  {journal} {Physical Review A}\ }\textbf
  {\bibinfo {volume} {82}},\ \bibinfo {pages} {13635} (\bibinfo {year}
  {2010})}\BibitemShut {NoStop}%
\bibitem [{\citenamefont {Duine}\ \emph {et~al.}(2011)\citenamefont {Duine},
  \citenamefont {Polini}, \citenamefont {Raoux}, \citenamefont {Stoof},\ and\
  \citenamefont {Vignale}}]{Duine2011}%
  \BibitemOpen
  \bibfield  {author} {\bibinfo {author} {\bibfnamefont {R.~A.}\ \bibnamefont
  {Duine}}, \bibinfo {author} {\bibfnamefont {M.}~\bibnamefont {Polini}},
  \bibinfo {author} {\bibfnamefont {A.}~\bibnamefont {Raoux}}, \bibinfo
  {author} {\bibfnamefont {H.~T.~C.}\ \bibnamefont {Stoof}}, \ and\ \bibinfo
  {author} {\bibfnamefont {G.}~\bibnamefont {Vignale}},\ }\href
  {http://stacks.iop.org/1367-2630/13/i=4/a=045010} {\bibfield  {journal}
  {\bibinfo  {journal} {New Journal of Physics}\ }\textbf {\bibinfo {volume}
  {13}},\ \bibinfo {pages} {45010} (\bibinfo {year} {2011})}\BibitemShut
  {NoStop}%
\bibitem [{\citenamefont {Wong}\ \emph {et~al.}(2012)\citenamefont {Wong},
  \citenamefont {Stoof},\ and\ \citenamefont {Duine}}]{Wong2012}%
  \BibitemOpen
  \bibfield  {author} {\bibinfo {author} {\bibfnamefont {C.~H.}\ \bibnamefont
  {Wong}}, \bibinfo {author} {\bibfnamefont {H.~T.~C.}\ \bibnamefont {Stoof}},
  \ and\ \bibinfo {author} {\bibfnamefont {R.~A.}\ \bibnamefont {Duine}},\
  }\href {http://link.aps.org/doi/10.1103/PhysRevA.85.063613} {\bibfield
  {journal} {\bibinfo  {journal} {Physical Review A}\ }\textbf {\bibinfo
  {volume} {85}},\ \bibinfo {pages} {63613} (\bibinfo {year}
  {2012})}\BibitemShut {NoStop}%
\bibitem [{\citenamefont {Bruun}(2011)}]{Bruun2011}%
  \BibitemOpen
  \bibfield  {author} {\bibinfo {author} {\bibfnamefont {G.~M.}\ \bibnamefont
  {Bruun}},\ }\href {http://stacks.iop.org/1367-2630/13/i=3/a=035005}
  {\bibfield  {journal} {\bibinfo  {journal} {New Journal of Physics}\ }\textbf
  {\bibinfo {volume} {13}},\ \bibinfo {pages} {35005} (\bibinfo {year}
  {2011})}\BibitemShut {NoStop}%
\bibitem [{\citenamefont {Sommer}\ \emph {et~al.}(2011)\citenamefont {Sommer},
  \citenamefont {Ku}, \citenamefont {Roati},\ and\ \citenamefont
  {Zwierlein}}]{Sommer2011}%
  \BibitemOpen
  \bibfield  {author} {\bibinfo {author} {\bibfnamefont {A.}~\bibnamefont
  {Sommer}}, \bibinfo {author} {\bibfnamefont {M.}~\bibnamefont {Ku}}, \bibinfo
  {author} {\bibfnamefont {G.}~\bibnamefont {Roati}}, \ and\ \bibinfo {author}
  {\bibfnamefont {M.~W.}\ \bibnamefont {Zwierlein}},\ }\href {\doibase
  http://www.nature.com/nature/journal/v472/n7342/abs/10.1038-nature09989-unlocked.html#supplementary-information}
  {\bibfield  {journal} {\bibinfo  {journal} {Nature}\ }\textbf {\bibinfo
  {volume} {472}},\ \bibinfo {pages} {201} (\bibinfo {year}
  {2011})}\BibitemShut {NoStop}%
\bibitem [{\citenamefont {Koschorreck}\ \emph {et~al.}(2013)\citenamefont
  {Koschorreck}, \citenamefont {Pertot}, \citenamefont {Vogt},\ and\
  \citenamefont {Kohl}}]{Koschorreck2013}%
  \BibitemOpen
  \bibfield  {author} {\bibinfo {author} {\bibfnamefont {M.}~\bibnamefont
  {Koschorreck}}, \bibinfo {author} {\bibfnamefont {D.}~\bibnamefont {Pertot}},
  \bibinfo {author} {\bibfnamefont {E.}~\bibnamefont {Vogt}}, \ and\ \bibinfo
  {author} {\bibfnamefont {M.}~\bibnamefont {Kohl}},\ }\href {\doibase
  10.1038/nphys2637} {\bibfield  {journal} {\bibinfo  {journal} {Nat Phys}\
  }\textbf {\bibinfo {volume} {9}},\ \bibinfo {pages} {405} (\bibinfo {year}
  {2013})}\BibitemShut {NoStop}%
\bibitem [{\citenamefont {Ariel}\ \emph {et~al.}(2011)\citenamefont {Ariel},
  \citenamefont {Mark},\ and\ \citenamefont {Martin}}]{Ariel2011}%
  \BibitemOpen
  \bibfield  {author} {\bibinfo {author} {\bibfnamefont {S.}~\bibnamefont
  {Ariel}}, \bibinfo {author} {\bibfnamefont {K.}~\bibnamefont {Mark}}, \ and\
  \bibinfo {author} {\bibfnamefont {W.~Z.}\ \bibnamefont {Martin}},\ }\href
  {http://stacks.iop.org/1367-2630/13/i=5/a=055009} {\bibfield  {journal}
  {\bibinfo  {journal} {New Journal of Physics}\ }\textbf {\bibinfo {volume}
  {13}},\ \bibinfo {pages} {55009} (\bibinfo {year} {2011})}\BibitemShut
  {NoStop}%
\bibitem [{\citenamefont {Mukherjee}\ \emph {et~al.}(2017)\citenamefont
  {Mukherjee}, \citenamefont {Yan}, \citenamefont {Patel}, \citenamefont
  {Hadzibabic}, \citenamefont {Yefsah}, \citenamefont {Struck},\ and\
  \citenamefont {Zwierlein}}]{Mukherjee2017}%
  \BibitemOpen
  \bibfield  {author} {\bibinfo {author} {\bibfnamefont {B.}~\bibnamefont
  {Mukherjee}}, \bibinfo {author} {\bibfnamefont {Z.}~\bibnamefont {Yan}},
  \bibinfo {author} {\bibfnamefont {P.~B.}\ \bibnamefont {Patel}}, \bibinfo
  {author} {\bibfnamefont {Z.}~\bibnamefont {Hadzibabic}}, \bibinfo {author}
  {\bibfnamefont {T.}~\bibnamefont {Yefsah}}, \bibinfo {author} {\bibfnamefont
  {J.}~\bibnamefont {Struck}}, \ and\ \bibinfo {author} {\bibfnamefont {M.~W.}\
  \bibnamefont {Zwierlein}},\ }\href@noop {} {\bibfield  {journal} {\bibinfo
  {journal} {Phys Rev Lett}\ }\textbf {\bibinfo {volume} {118}},\ \bibinfo
  {pages} {123401} (\bibinfo {year} {2017})}\BibitemShut {NoStop}%
\bibitem [{\citenamefont {Chiacchiera}\ \emph {et~al.}(2009)\citenamefont
  {Chiacchiera}, \citenamefont {Lepers}, \citenamefont {Davesne},\ and\
  \citenamefont {Urban}}]{Chiacchiera2009}%
  \BibitemOpen
  \bibfield  {author} {\bibinfo {author} {\bibfnamefont {S.}~\bibnamefont
  {Chiacchiera}}, \bibinfo {author} {\bibfnamefont {T.}~\bibnamefont {Lepers}},
  \bibinfo {author} {\bibfnamefont {D.}~\bibnamefont {Davesne}}, \ and\
  \bibinfo {author} {\bibfnamefont {M.}~\bibnamefont {Urban}},\ }\href
  {http://link.aps.org/doi/10.1103/PhysRevA.79.033613} {\bibfield  {journal}
  {\bibinfo  {journal} {Physical Review A}\ }\textbf {\bibinfo {volume} {79}},\
  \bibinfo {pages} {33613} (\bibinfo {year} {2009})}\BibitemShut {NoStop}%
\bibitem [{\citenamefont {Ueda}(2010)}]{Ueda2010}%
  \BibitemOpen
  \bibfield  {author} {\bibinfo {author} {\bibfnamefont {M.}~\bibnamefont
  {Ueda}},\ }\href@noop {} {\emph {\bibinfo {title} {Fundamentals and New
  Frontiers of Bose--Einstein Condensation}}}\ (\bibinfo  {publisher} {World
  Scientific Printers},\ \bibinfo {year} {2010})\BibitemShut {NoStop}%
\bibitem [{\citenamefont {Jo}\ \emph {et~al.}(2009)\citenamefont {Jo},
  \citenamefont {Lee}, \citenamefont {Choi}, \citenamefont {Christensen},
  \citenamefont {Kim}, \citenamefont {Thywissen}, \citenamefont {Pritchard},\
  and\ \citenamefont {Ketterle}}]{Jo2009}%
  \BibitemOpen
  \bibfield  {author} {\bibinfo {author} {\bibfnamefont {G.-B.}\ \bibnamefont
  {Jo}}, \bibinfo {author} {\bibfnamefont {Y.-R.}\ \bibnamefont {Lee}},
  \bibinfo {author} {\bibfnamefont {J.-H.}\ \bibnamefont {Choi}}, \bibinfo
  {author} {\bibfnamefont {C.~A.}\ \bibnamefont {Christensen}}, \bibinfo
  {author} {\bibfnamefont {T.~H.}\ \bibnamefont {Kim}}, \bibinfo {author}
  {\bibfnamefont {J.~H.}\ \bibnamefont {Thywissen}}, \bibinfo {author}
  {\bibfnamefont {D.~E.}\ \bibnamefont {Pritchard}}, \ and\ \bibinfo {author}
  {\bibfnamefont {W.}~\bibnamefont {Ketterle}},\ }\href@noop {} {\bibfield
  {journal} {\bibinfo  {journal} {Science}\ }\textbf {\bibinfo {volume}
  {325}},\ \bibinfo {pages} {1521} (\bibinfo {year} {2009})}\BibitemShut
  {NoStop}%
\bibitem [{\citenamefont {Zhang}\ and\ \citenamefont {Ho}(2011)}]{Zhang2011}%
  \BibitemOpen
  \bibfield  {author} {\bibinfo {author} {\bibfnamefont {S.}~\bibnamefont
  {Zhang}}\ and\ \bibinfo {author} {\bibfnamefont {T.-L.}\ \bibnamefont {Ho}},\
  }\href@noop {} {\bibfield  {journal} {\bibinfo  {journal} {New Journal of
  Physics}\ }\textbf {\bibinfo {volume} {13}},\ \bibinfo {pages} {055003}
  (\bibinfo {year} {2011})}\BibitemShut {NoStop}%
\bibitem [{\citenamefont {Sanner}\ \emph {et~al.}(2012)\citenamefont {Sanner},
  \citenamefont {Su}, \citenamefont {Huang}, \citenamefont {Keshet},
  \citenamefont {Gillen},\ and\ \citenamefont {Ketterle}}]{Sanner2012}%
  \BibitemOpen
  \bibfield  {author} {\bibinfo {author} {\bibfnamefont {C.}~\bibnamefont
  {Sanner}}, \bibinfo {author} {\bibfnamefont {E.~J.}\ \bibnamefont {Su}},
  \bibinfo {author} {\bibfnamefont {W.}~\bibnamefont {Huang}}, \bibinfo
  {author} {\bibfnamefont {A.}~\bibnamefont {Keshet}}, \bibinfo {author}
  {\bibfnamefont {J.}~\bibnamefont {Gillen}}, \ and\ \bibinfo {author}
  {\bibfnamefont {W.}~\bibnamefont {Ketterle}},\ }\href@noop {} {\bibfield
  {journal} {\bibinfo  {journal} {Phys Rev Lett}\ }\textbf {\bibinfo {volume}
  {108}},\ \bibinfo {pages} {240404} (\bibinfo {year} {2012})}\BibitemShut
  {NoStop}%
\end{thebibliography}%

\end{document}